\documentclass[12pt,a4]{article}

\usepackage{amsthm,amsmath,latexsym,amssymb,amsfonts,amscd}
\usepackage{graphics,lscape,fancyhdr,array,stmaryrd,euscript}
\usepackage{empheq}
\usepackage{dsfont}
\usepackage{verbatim}
\usepackage{color,tikz,tikz-cd}
\usetikzlibrary[snakes]
\usepackage{relsize,slashed}
\numberwithin{equation}{section}
\usepackage{hyperref}

\newcommand{\bep}{\begin{picture}}
\newcommand{\eep}{\end{picture}}

\newcounter{YoungHeight}\newcounter{YoungWidth}

\newcounter{Mul1}\newcounter{Mul2}\newcounter{Mul3}\newcounter{Mul4}
\newcounter{A0}\newcounter{A1}\newcounter{A2}
\newcounter{B3}
\newcounter{C3}\newcounter{C4}
\newcounter{D1}\newcounter{D2}\newcounter{D3}
\newcounter{T0}\newcounter{T1}

\newlength{\txtHShift}

\newlength{\txtWidth}

\newcommand{\HalfLength}[2]{\setcounter{Mul1}{#1}\setcounter{Mul2}{#1}\addtocounter{Mul1}{\value{Mul2}}\addtocounter{Mul1}{\value{Mul2}}%
\addtocounter{Mul1}{\value{Mul2}}\addtocounter{Mul1}{\value{Mul2}}\setcounter{#2}{\value{Mul1}}}

\newcommand{\Add}[3]{\setcounter{#1}{#2}\addtocounter{#1}{#3}}

\newcommand{\Length}[1]{#10}
\newcommand{\YoungScale}{}

\newcommand{\shiftedText}[2]{{\hspace{#1}#2}}
\newcommand{\calcHShift}[1]{\settowidth{\txtWidth}{#1}\setlength{\txtHShift}{-0.5\txtWidth}}

\newcommand{\TextTop}[3]{{\calcHShift{#1}\HalfLength{#2}{T0}\Add{T1}{\Length{#3}}{-9}\put(\value{T0},\value{T1}){\shiftedText{\txtHShift}{#1}}}}

\newcommand{\RectT}[3]{\bep(\Length{#1},\Length{#2})\put(0,0){\line(1,0){\Length{#1}}}\put(0,0){\line(0,1){\Length{#2}}}%
\put(\Length{#1},\Length{#2}){\line(-1,0){\Length{#1}}}\put(\Length{#1},\Length{#2}){\line(0,-1){\Length{#2}}}#3{#1}{#2}\eep}

\newcommand{\RectBRow}[4]{{\bep(\Length{#1},20)\put(0,0){\RectT{#2}{1}{\TextTop{#4}}}%
\put(0,10){\RectT{#1}{1}{\TextTop{#3}}}\eep}}

\newcommand{\BlockApar}[2]{\parbox{\Length{#1}pt}{\YoungScale\bep(\Length{#1},\Length{#2}){\Add{A1}{#1}{1}\Add{A2}{#2}{1}}%
\multiput(0,0)(10,0){\value{A1}}{\line(0,1){\Length{#2}}}\multiput(0,0)(0,10){\value{A2}}{\line(1,0){\Length{#1}}}%
\setcounter{YoungHeight}{\Length{#2}}\setcounter{YoungWidth}{\Length{#1}}\eep}}

\newcommand{\YoungpA}{\BlockApar{1}{1}}
\newcommand{\YoungpB}{\BlockApar{2}{1}}

\newcommand{\YoungpAA}{\BlockApar{1}{2}}

\usepackage{setspace}

 \usepackage[numbers,sort&compress]{natbib}
 \usepackage[nottoc]{tocbibind}
\newcommand{\pl}{\partial}
\newcommand{\besubeqs}{\begin{subequations}}
\newcommand{\esubeqs}{\end{subequations}}

\newcommand{\fdu}[2]{{}_{#1}{}^{#2}\,}

\newcommand{\Tr}{{\mathrm{Tr}\,}}

\newcommand{\PPP}{{\boldsymbol{P}}}
\newcommand{\LLL}{{\boldsymbol{L}}}
\newcommand{\KKK}{{\boldsymbol{K}}}
\newcommand{\DDD}{{\boldsymbol{D}}}
\newcommand{\TTT}{{\boldsymbol{T}}}

\newcommand{\ttt}{{\boldsymbol{t}}}
\newcommand{\SSS}{{\boldsymbol{S}}}
\newcommand{\UUU}{{\boldsymbol{U}}}
\newcommand{\RRR}{{\boldsymbol{R}}}
\newcommand{\WWW}{{\boldsymbol{W}}}
\newcommand{\VVV}{{\boldsymbol{V}}}
\newcommand{\sss}{{\boldsymbol{s}}}
\newcommand{\hs}{{\mathfrak{hs}}}
\newcommand{\SP}{{\boldsymbol{SP}}}
\newcommand{\SK}{{\boldsymbol{SP}}}

\newcommand{\SSDD}{{\boldsymbol{SD}}}
\newcommand{\SSLL}{{\boldsymbol{SL}}}

\newcommand{\mak}{\mathcal{K}}

\begin{document}
\hfill
\vspace{-1.5cm}
\vskip 0.05\textheight
\begin{center}
{\Large\bfseries 
Holography of New Conformal higher-spin Gravities in $\boldsymbol{3d}$}

\vspace{0.4cm}

\vskip 0.03\textheight

 Iva \textsc{Lovrekovic}${}^a$ 

\vskip 0.03\textheight

\vspace{5pt}
{\em
$^a$ Institute for Theoretical Physics, TU Wien, \\
 Wiedner Hauptstr. 8-10, 1040 Vienna, Austria\\

}
\end{center}

\vskip 0.02\textheight

\begin{abstract}

We study the holography of new conformal higher-spin theories with general and near-horizon boundary conditions. General boundary conditions lead to an ASA that is the loop algebra of the underlying gauge algebra. Near-horizon boundary conditions give rise to u(1) currents at the boundary. For each set of boundary conditions, we consider the conformal graviton alone and its combinations with a vector field and a spin-3 field. The near-horizon boundary conditions lead to three-dimensional Bañados–Teitelboim–Zanelli (BTZ)-like black holes, which can have up to four horizons, Lobachevsky solutions, and generalizations of the BTZ-like solution. We verify these results independently by deriving them from the Cotton tensor.
\end{abstract}

\newpage
\tableofcontents

\section{Introduction}

The theory of conformal higher spins (CHS) has been known for a long time \cite{Fradkin:1985am}. These original CHS theories describe conformal symmetric traceless fields with arbitrary spin and a single derivative in the linearized gauge transformations of the respective conformal fields \cite{Pope:1989vj,Fradkin:1989xt}. The number of such fields in these theories is usually infinite. 


Three-dimensional higher-spin (HS) theories are typically based on the Chern-Simons (CS) action which relies on an underlying gauge algebra. 
The Chern-Simons action naturally accommodates an infinite tower of higher-spin fields. A consistent truncation of this infinite tower results in a finite set of interacting, non-propagating spin-s fields
s=2,3,…,N \cite{Campoleoni:2010zq, Campoleoni:2008jq}. In their work, Pope and Townsend \cite{Pope:1989vj} analyzed this infinite tower of higher-spin fields and apart from the construction they introduced, they predicted the existence of another construction featuring an infinite number of theories, each containing a finite number of higher-spin fields. This construction was used in the development of a new conformal higher-spin gravity theory (conformal HiSGRA) \cite{Grigoriev:2019xmp}.


The program of holographic analysis for three-dimensional higher-spin (HS) gauge theories began with \cite{Campoleoni:2010zq,Henneaux:2010xg}, which generalized the work of Brown and Henneaux \cite{Brown:1986nw} to massless higher spins \cite{Bergshoeff:1989ns,Blencowe:1988gj}. This analysis demonstrated that the two copies of the Virasoro algebra, with a central extension and a central charge dependent on the AdS radius, transform into two copies of the W-algebra, with the central charge matching that of pure gravity \cite{Campoleoni:2011hg}. 

Further steps in the holographic analysis of CS theory with the $sl(2)\oplus sl(2)$ (3D Einstein gravity) gauge group explored general boundary conditions (bcs) and near-horizon bcs. General boundary conditions in the CS formulation, as described in \cite{Grumiller:2016pqb}, lead to an asymptotic symmetry algebra that forms a loop algebra of the underlying gauge algebra. This result also holds for the new conformal HiSGRA theories in 3D.

The Near-horizon bcs \cite{Afshar:2016wfy} lead to a configuration known as the black flower. Its asymptotics include a number of affine 
$u(1)$ charges that commute with the Hamiltonian and correspond to soft hair \cite{Hawking:2016msc}. The higher-spin extension of these bcs consists of bosonic higher-spin fields coupled to gravity in $AdS_3$
  \cite{Blencowe:1988gj}. The near-horizon holography of conformal HiSGRA in 3D \cite{Grigoriev:2019xmp} similarly yields charges associated with soft hair.

New conformal HiSGRA in 3D has been studied so far from the perspective of a non-relativistic and ultra-relativistic limit, in which the spectra of the theory and their comparison with the known Galilean and Carroll theories in three dimensions were studied \cite{Lovrekovic:2021dvi,Lovrekovic:2022lwv}. The work on Carroll limit also considers the holographic aspect of the theory and specific metric solutions which indicate non-vanishing entropy.
Here, we comment on the most general bcs and study near-horizon bcs of conformal HiSGRA in three dimensions.
Main results of this paper are: 
(i) Imposing near-horizon bcs leads to a set of canonical charges generating an affine $\mathfrak{u}(1)$ algebra. This structure holds for higher-spin algebras in the tower of new conformal HiSGRA theories and arises from the boundary conditions defined along the diagonal generators in the chosen representation, and linear dependence on the state dependent functions.
(ii) Using the Chern–Simons holonomy method, we show that the horizon entropy depends only on the zero modes of the affine currents. Its overall normalization is sensitive to the embedding of $\mathfrak{so}(3,2)$ into the higher-spin algebra. Inequivalent embeddings imply distinct gravitational sectors (fields of different spin) despite  near-horizon algebras built from $U(1)$-currents.
(iii) We reconstruct representative metrics for embeddings in $\mathfrak{so}(3,2)$, $\mathfrak{sl}(4)$, and $\mathfrak{sl}(5)$. 
Embedding choice affects geometric features even when the underlying near-horizon algebra remains similar.

First, in Section 2, we review the new conformal HiSGRA theories. Experts in higher-spin gravity can skip this section and proceed directly to Section 3.
In Section 3, we review the holography of CS theory and comment on general bcs with a fixed chemical potential. 
In Section 4, we focus on near-horizon bcs for spin-s fields in the first three new conformal HiSGRA theories, considering CG theory, CG with a vector field, and CG with a spin-3 field. For these cases, we construct a map between the generators in the conformal "PKLD" basis of the gauge algebra and those in the "WL" fundamental representation. This basis simplifies the implementation of near-horizon bcs.

When expressed in the metric formulation, these bcs yield generalizations of well-known solutions in 3D gravity. For the rotating black hole in 3D, we provide a generalization for 3D CG and comment on higher-spin extensions.

 Section 5 summarizes and concludes our results.


\section{ New Conformal higher-spin Theories}
\label{sec:newchs}
The construction of the new conformal HiSGRA is based on the Chern-Simons action.
In the standard Chern-Simons action
\begin{align}
    S[\omega]&=\int \Tr\left[\omega \wedge d \omega +\frac23 \omega\wedge \omega\wedge \omega\right]\, \label{cs}
\end{align}
 the field $\omega$ is a Lie-algebra-valued one-form. Which means that $\omega$ is written as an expansion in terms of the generators of the chosen Lie-algebra. When the Lie algebra is $so(3,2)$, the action is equivalent to the CG action for $\omega$ valued in Lorentz algebra \cite{Horne:1988jf}. The action (\ref{cs}) based on $so(3,2)$  admits a higher-spin extension \cite{Pope:1989vj} that leads to a theory with an infinite tower of higher-spin fields. The equations of motion are \begin{align} F=d\omega+\omega\wedge\omega=0.\label{eomcs}\end{align}  Based on the (\ref{cs}) action and $so(3,2)$ embedding algebra, (as predicted in \cite{Pope:1989vj}) one can define an infinite number of theories, new conformal HiSGRA, each with a finite number of higher-spin fields \cite{Grigoriev:2019xmp}. These higher-spin fields consist of Fradkin-Tseytlin fields and conformal partially massless fields. Fradkin-Tseytlin fields are symmetric traceless tensors $\phi_{a_1...a_s}$ with linearized gauge transformations 
\begin{align}
\delta\phi_{a_1...a_s}=\partial_{a_1}\xi_{a_2...a_s}+\text{permutations}-\text{traces}.
\end{align}
By conformal partially massless fields we mean the symmetric traceless tensors that have higher derivatives in their gauge transformations
\begin{align}
\delta\phi_{a_1...a_s}=\partial_{a_1}...\partial_{a_t}\xi_{a_{t+1}...a_s}+\text{permutations}-\text{traces}.
\end{align}
Conformal partially-massless fields have been first introduced in \cite{Bekaert:2013zya} where they have been referred to as ``higher-depth Fradkin-Tseytlin fields".
The latter can also be understood as boundary values for the real partially massless fields in higher-dimensional AdS space. 
These kinds of fields are going to appear as a result of gauging the higher-spin algebras constructed below. We are going to need a higher-spin extension of $so(3,2)$, i.e. algebras that have $so(3,2)$ as a subalgebra and also contain various other nontrivial representations of $so(3,2)$.

\subsection{Two Simple higher-spin Algebras}
\label{sec:hsa}

\paragraph{Finite-dimensional algebras. } Algebras with a finite spectrum of Fradkin-Tseytlin fields are constructed by taking any nontrivial  finite-dimensional irreducible representation $V$ of the $so(3,2)$ algebra, for example an irreducible tensor or a spin-tensor. After that, we evaluate $U(so(d,2))$ in $V$, i.e. multiply the generators\footnote{Indices $A,B,...=0,...,4$ are the indices of the $so(3,2)$ with the invariant metric denoted $\eta_{AB}$.} $\TTT_{AB}=-\TTT_{BA}$ of the $so(3,2)$ in this representation and identify the algebra that they generate, which we denote with $\hs(V)$. That algebra is associative by construction.
Since $V$ is irreducible, the algebra is  the algebra of all matrices of size $\dim V$, i.e. $\mathrm{End}(V)$. Therefore $\hs(V)=\mathrm{End}(V)$ comes equipped with a specific embedding of $so(3,2)$ generators via $\TTT_{AB}$ by construction. This algebra can be understood as the quotient $U(so(3,2))/I$ by a two-sided ideal $I$, that is the annihilator of $V$, i.e. $I$ contains all polynomials in $\TTT_{AB}$ that vanish on $V$. The next step is decomposition of $\hs(V)$ into irreducible $so(3,2)$-modules --- this decomposition, as is shown below, determines the spectrum of Fradkin-Tseytlin fields. We are interested in the holography of the action (\ref{cs}).

To summarize, there are infinitely many higher-spin extensions of $so(3,2)$ that are labelled by various irreducible $so(3,2)$-modules:\footnote{Various peculiarities of higher-derivative theories of type $\square^k\Phi(x)=0$ were noticed in \cite{Brust:2016gjy}. It would be interesting to see if the finite-dimensional higher-spin algebras can be explained from the CFT point of view. }
\begin{align}
    \hs(V) =\mathrm{End}(V)=V\otimes V^*\,.
\end{align}
Spin-tensor representations $V$ are also allowed.\footnote{For example, a toy-model was studied in \cite{Pope:1989vj}, where $\hs$ was taken to be $so(4,2)$. This can be understood as $sl(V)$ for $V$ being the spinor representation of $so(3,2)$. }
Further extensions can be obtained by tensoring $\hs(V)$ with the matrix algebra $\mathrm{Mat}_N$, which turns the otherwise abelian spin-one field associated with $1$ of $\mathrm{End}(V)$ into the Yang-Mills field.\footnote{One can also project onto the symmetric and anti-symmetric parts of the tensor product $V\otimes V$ (as a matter of fact $V\sim V^*$ for finite-dimensional representations of $so(3,2)$), the only restriction being that it to contain the adjoint, i.e. $so(3,2)$ itself. In this way one can define higher-spin analogs of $so(V)$ and $sp(V)$. } Note, that $\hs(V)=gl(V)=sl(V)\oplus u(1)$ as a Lie algebra and the spin-one field always decouples unless the Yang-Mills groups are added. In other words, these are not infinite-dimensional higher-spin algebras but matrix realizations with higher-spin interpretation via so(3,2) decomposition.
Since the finite-dimensional higher-spin algebras result from $\mathrm{End}(V)$ they are equipped with a canonical trace operation $\Tr$, which allows us to write down the CS action.  

\paragraph{Example 1.} As the simplest example, let us take the vector representation, which is denoted by one-cell Young diagram $\YoungpA$. The corresponding algebra, $\hs(\YoungpA)$, has the following spectrum:\footnote{When treated as a higher-spin algebra in $AdS_4$ this spectrum corresponds to a spin-2 field, partially-massless depth-$3$ spin-3 field and to a spin-one field.}
\begin{align}
    \hs(\YoungpA)&=\YoungpA\otimes \YoungpA= \YoungpAA \oplus \YoungpB\oplus \bullet \,.\label{hsconstruction}
\end{align}
The algebra is just the matrix algebra with $(d+2)^2$ generators $\ttt\fdu{A}{B}$ decomposed with respect to $so(d,2)$ (we keep the discussion $d$-dimensional here, otherwise, $d=3$). The $gl_{d+2}$ commutation relations are
\begin{align}
    [\ttt\fdu{A}{B}, \ttt\fdu{C}{D}]&=-\delta\fdu{A}{D}\ttt\fdu{C}{B} +\delta\fdu{C}{B} \ttt\fdu{A}{D}\,. \label{alg2}
\end{align}
In the $so(d,2)$ base, the irreducible generators are $\TTT_{AB}=-\TTT_{BA}$, $\SSS_{AB}=\SSS_{BA}$ and $\RRR$, i.e.
\begin{align}
    \TTT_{AB}&=\ttt_{A|B}-\ttt_{B|A}\,, & \SSS_{AB}&=\ttt_{A|B}+\ttt_{B|A}-\frac{2}{d+2}\ttt\fdu{C}{C}\eta_{AB}\,, &&\RRR=\ttt\fdu{C}{C}\,. \label{eq15}
\end{align}
The commutation relations read 
\besubeqs
\begin{align}
    [\TTT_{AB},\TTT_{CD}]&= \eta_{BC} \TTT_{AD}-\eta_{AC} \TTT_{BD}-\eta_{BD} \TTT_{AC}+\eta_{AD} \TTT_{BC}\,,\\
    [\TTT_{AB},\SSS_{CD}]&= \eta_{BC}\SSS_{AD}-\eta_{AC}\SSS_{BD}+\eta_{BD}\SSS_{AC}-\eta_{AD}\SSS_{BC}\,,\\
    [\SSS_{AB},\SSS_{CD}]&= \eta_{BC}\TTT_{AD}+\eta_{AC}\TTT_{BD}+\eta_{BD}\TTT_{AC}+\eta_{AD}\TTT_{BC}\,,
\end{align}
\esubeqs
while $\RRR$ commutes with everything since it is associated with $1$ in $gl(V)$. We can truncate it away. As a result our theory contains two fields:
\begin{align}
    \omega&= \omega^{A,B} \TTT_{AB} +\omega^{AB} \SSS_{AB}
\end{align}
The first one is the conformal graviton and the second one is the similar to the partially massless spin-3 case. It is studied in the paper. In flat space background we find a metric-like field with the following gauge symmetries
\begin{align}\label{spinthreeg}
    \delta\phi^{abc}&= \pl^a\pl^b\pl^c \xi -\frac15(\eta^{ab}\pl^c \square \xi+\eta^{ac}\pl^b \square \xi+\eta^{ac}\pl^b \square \xi)\,.
\end{align}
The bilinear form, (denoted by $\langle A,B \rangle\equiv \Tr(AB)$ for some elements $A,B$ of the corresponding algebra) is given by 
\begin{align}
    \langle\ttt\fdu{A}{B}, \ttt\fdu{C}{D}\rangle&=\delta\fdu{A}{D}\delta\fdu{C}{B}
\end{align}
from which we easily get for $so(d,2)$
\begin{align}
    \langle \TTT_{A,B}, \TTT_{C,D}\rangle&=2(\eta_{AD}\eta_{CB}-\eta_{BD}\eta_{CA}) \label{eq210}\\
    \langle \SSS_{A,B}, \SSS_{C,D}\rangle&=2(\eta_{AD}\eta_{CB}+\eta_{BD}\eta_{CA}-\frac{2}{d+2}\eta_{AB}\eta_{CD})\label{eq211}
\end{align}

\paragraph{Example 2.} We can take $V$ to be the spinor of $so(3,2)$ then the resulting algebra is nothing but $so(4,2)$. Let us simply write its commutation relations
\besubeqs
\begin{align}
    [\TTT_{AB},\TTT_{CD}]&= \eta_{BC} \TTT_{AD}-\eta_{AC} \TTT_{BD}-\eta_{BD} \TTT_{AC}+\eta_{AD} \TTT_{BC}\,,\\
    [\TTT_{AB},\UUU_{C}]&= \eta_{BC}\UUU_{A}-\eta_{AC}\UUU_{B}\,,\label{pkldveb}\\
    [\UUU_{A},\UUU_{C}]&= \TTT_{AC} \label{pkldvec}
\end{align}\label{pkldv}
\esubeqs
So the gauge field is
\begin{align}
    \omega&= \omega^{A,B} \TTT_{AB} +\omega^{A} \UUU_{A}\label{eq110}
\end{align}
and the second field describes a conformal analog of the partially-massless spin-2 field:
\begin{align}\label{spintwog}
    \delta \phi_{ab}=\pl_a \pl_b\xi -\frac{1}{3}\eta_{ab} \square \xi\,.
\end{align}

When we represent the generators of this basis, we refer to to corresponding matrices as written in PKLD basis. After split in the light cone coordinates indices are denoted with $(+,-,a)$ where $a=1,..,3$. The name comes from the common use of letters $\PPP_a,\KKK_a,\LLL_{ab},\DDD$, for  $t_{a+}\equiv\PPP_a,t_{ab}\equiv\LLL_{ab},t_{a-}\equiv\KKK_{a},t_{+-}\equiv-\DDD$  to refer to generators of translations, special conformal transformations, Lorentz rotations and dilatations respectively, which appear in each of the new conformal HiSGRA theories that we consider.

\section{General boundary conditions}

We outline how to impose boundary conditions in 3D CS theory while considering the most general boundary conditions.

\subsection{A short review of Chern–Simons holography}
First we summarize the standard holographic treatment of a
three-dimensional Chern--Simons theory in the vielbein formulation. The framework in which we are working on follows the methods of \cite{Banados:1994tn, Banados:1998ta} and is analogous to the framework summarized in \cite{Campoleoni:2010zq}.
Writing the theory on a manifold with topology $\Sigma\times\mathbb{R}$,
the action is put in canonical $2+1$ form as
\[
 I=\frac{k}{4\pi}\int_{\mathbb{R}}dt\int_{\Sigma}d^2x\,
 \epsilon^{ij}g_{\mu\nu}
 \left(\dot{\omega}^{\mu}_{i}\omega^{\nu}_{j}
 +\omega^{\mu}_{t}F^{\nu}_{ij}\right)+B(\partial\Sigma).
\]
Here $B(\partial\Sigma)$ is a boundary term chosen so that the variational
principle and gauge invariance are well-defined for the imposed boundary
conditions. The temporal component $\omega_t^\mu$ acts as a Lagrange
multiplier imposing the flatness constraints
\[
 G_\mu =\frac{k}{4\pi} g_{\mu\nu}\epsilon^{ij}F^\nu_{ij},
\]
which generate gauge transformations on phase space. To make these
generators functionally differentiable in the presence of a boundary, one
adds canonical boundary charges $Q(\xi)$, whose variation is
\[
 \delta Q[\xi]=-\frac{k}{2\pi}\oint_{\partial\Sigma}
 \langle \xi\,\delta\omega\rangle .
\]
The algebra of the improved generators then acquires the
standard boundary contribution,
\[
 \{Q(\lambda),Q(\xi)\}^{*}
 =Q([\lambda,\xi])+K(\xi,\lambda),
\]
where $K$ is the possible central extension. Then, one can fix the
radial gauge,
\[
 \omega_\rho=b^{-1}(\rho)\partial_\rho b(\rho),
 \qquad
 \omega=b^{-1}(\rho)\bigl(d+\Omega(t,\varphi)\bigr)b(\rho),
\]
which separates the radial dependence from the boundary data. This form is
the starting point for the analysis of both general and near-horizon
boundary conditions\footnote{Slightly broader review is given in the Appendix A}.


\subsection{Imposing the general boundary conditions}
As we have seen in the previous section,
the CS action is in this approach usually put in the canonical form on the manifold with the topology $\Sigma \times \mathbb{R}$. 
Here $\Sigma$ is a two-dimensional manifold on the boundary and $\mathbb{R}$  denotes one-dimensional manifold. $\Sigma$ is usually considered as a disc with coordinates $\rho$ and $\phi$. The action can be written in the 2+1 decomposition.
To impose the general boundary conditions to the CS theory based on the gauge group G, first,  
we will partially fix to radial gauge. This means that
we can write the  field $\omega$ as \begin{align}
 \omega=b(\rho)^{-1}(d+\Omega(t,\phi))b(\rho) \label{bcsx1}
 \end{align} while group element $b$ depends on the radial coordinate.
 The particular choice of bcs is  done by choosing $\Omega(t,\phi)$, and when we want to translate the solution to metric formulation, it will depend on the precise form of the group element $b$.  
Here, the $\Omega(t,\phi)$ is defined by expanding it in all the generators of the gauge group, multiplied by arbitrary functions. We can write $\Omega(t,\phi)=\tau(t,\phi)_{A(k),B(l)} \TTT^{A(k),B(l)} d\phi+\mu(t,\phi)_{A(k),B(l)} \TTT^{A(k),B(l)} dt $ where  $\tau(t,\phi)^{A(k),B(l)}$ and $\mu(t,\phi)^{A(k),B(l)}$ are arbitrary functions. $A(k),B(l)$ denote the indices of the generators, with $k$ and $l$ going from 1 to $s_1$ and $s_2$ respectively. With total number of indices $s=s_1+s_2$. 
$\mu(t,\phi)^{A(k),B(l)}$ are the chemical potentials which we here set to zero, while the functions $\tau(t,\phi)^{A(k),B(l)}$ become charges and define asymptotic symmetry algebra. The fact that we used the expansion in all the generators and functions $\tau(t,\phi)$ dependent on $(t,\phi)$, and allowed these functions to vary, the boundary conditions can be referred to as most general. The chemical potentials are here set to zero for simplicity, as they play a role for consideration of thermodynamics of the solutions in the system. In this work we consider specific solutions for the near horizon boundary conditions (below) where the chemical potentials are not set to zero.  
The corresponding gauge parameter $\xi$ is also expanded in the generators $\TTT^{A(k),B(l)}$ of the group. The gauge parameter is restricted by considering the transformations \begin{align}
\delta_{\xi}\omega=d\xi+[\omega,\xi]\label{gaugetrafo}
\end{align}
 that conserve the boundary conditions.

In our picture of CS theory, beside gauge transformations, there is an invariance under diffeomorphisms, and other symmetries depending on the precise theory in the tower of conformal HiSGRA theories. For example, for conformal graviton, there is an additional invariance under Lorentz transformations, special conformal transformations and dilatation. For the conformal graviton and higher-spin field, in addition there would be invariance under higher-spin symmetries. 
This allows that there is an independent set of conserved charges due to that symmetry.

We need to find the restrictions on the function $\xi$ that conserve the gauge transformations (\ref{gaugetrafo}).
 Then it is possible to determine the infinitesimal changes of the functions $\tau_{A(k),B(l)}$ under transformations that preserve bcs (\ref{bcsx1}). 
The variation of the canonical boundary charges which is  background independent \cite{Banados:1998ta,Banados:1994tn,Banados:1998gg,Carlip:2005zn} is equal to 
\begin{align}
    \delta Q[\xi]=-\frac{k}{2\pi}\oint \left<\xi \delta\omega\right>\label{varcharge}.
\end{align}
Functional integration of (\ref{varcharge}) gives a  well known result for the canonical boundary charges \cite{Brown:1986nw}\cite{Campoleoni:2010zq}
\begin{align}
Q[\xi]=-\frac{k}{2\pi}\oint d\phi \langle \xi\omega \rangle.
\end{align}

The algebra of the charges is given by
\begin{align}
    \{Q(\xi),Q(\zeta)\}=Q([\xi,\zeta])+Z(\xi,\zeta) \label{jcharge}
\end{align}
 where
 $\zeta$ is also a gauge parameter and $Z$ is a possible central term.
 Using the Fourier modes, one can write a charge as 
 \begin{align}
     Q(\xi)=\sum_n T_n\xi^n \label{eq315}
 \end{align}
 and the modes with \begin{align}
     \xi^n=\frac{1}{4\pi}\int d\phi \xi(\phi)e^{in\phi}\label{eq316}.
 \end{align}
The $T_n$ are the coefficients that are used to define the asymptotic symmetry algebra of the charges, defined through $\omega=\frac{1}{k}\sum_{m=-\infty}^{\infty}T_m e^{im\phi}$. 
They are the generators of the asymptotic symmetry algebra \cite{Banados:1994tn}.
 
To find the ASA these modes have to be inserted into (\ref{jcharge}). In the general case, for the full set of generators and without imposing any special conditions, this leads to the loop algebra of the underlying gauge algebra. 
Since the boundary conditions allow all the fields to vary, the gauge parameters have no imposed conditions from the allowed gauge transformations. 

Due to the linear dependence of the gauge parameter on the fields in this construction, and boundary conditions that allow the functions along all the generators to vary, asymptotic symmetry algebra becomes a loop algebra of the underlying gauge algebra. Therefore, if we fix the chemical potentials, and do not impose further constraints,
the number of charges obtained is going to be equal to the number of the gauge generators. 

In 3D conformal gravity (CG), these bcs lead to twenty independent functions. When the chemical potentials are set to zero, ten functions remain, which represent charges. Adding a vector field to CG and imposing general bcs results in 15 charges. Adding a spin-3 field to CG and imposing general bcs results in a total of 24 charges. Therefore, the number of underlying gauge generators will tell us the number of corresponding independent charges that will remain after considering general bcs and zero chemical potentials, in new conformal HiSGRA in 3D.

To understand in what sense the boundary conditions are the most general,
one should recall the definition of the theory in terms of the Chern--Simons
action. As described in Section~2, the theory is written in terms of the
connection $\omega$, which is a Lie-algebra-valued one-form, i.e. it is
expanded in the generators of the underlying gauge algebra. In the
asymptotic region, one can impose boundary conditions either by retaining
all generators or by setting some of them to zero from the outset. Here, we
keep the entire set of generators. We will see how this differs from the
near-horizon boundary conditions, where only a subset of generators is
retained.

For these bcs, by choosing the appropriate group element, one can obtain any of the particular metrics that were originally derived under a stricter set of boundary conditions (bcs) \cite{Grumiller:2016kcp}. When the boundary conditions are already restricted, the described procedure yields a specific algebra of charges. This situation arises while imposing near-horizon boundary conditions.

\section{Near-horizon boundary conditions}
The name near-horizon boundary conditions is inspired by the fact that they lead to black hole solutions when the underlying gauge group of the CS theory is $sl(N,\mathbb{R})\oplus sl(N,\mathbb{R})$, represented in the WL representation\footnote{WL representation we call the representation in which the generators $W_m^s$ (for $-s+1\leq m \leq s-1$, and s, spin) are used as a basis. The generator $W_{s-1}^s=L_1^{s-1}$, while the higher generators are defined recursively $W^s_{m-1}=-\frac{1}{m+s-1}[L_{-1},W^s_m]$, until we get $W^s_{1-s}=L_{-1}^{s-1}$ \cite{Campoleoni:2024ced}.}

When the underlying gauge algebra of the theory is $sl(N,\mathbb{R})\oplus \overline{sl(N,\mathbb{R})}$, the field $\omega$ from (\ref{cs}) and $\overline{\omega}$  take value in the  generators $L_0$ and $W_0^{(s)}$ of Cartan subalgebra of the $sl(N,\mathbb{R})$ and $\overline{sl(N,\mathbb{R})}$ respectively. Here, we denoted the second $\overline{sl(N,\mathbb{R})}$ with an overline, analogously to field $\overline{\omega}$ that takes values in the generators of its algebra. 
These generators  have diagonal matrices in the  WL representation, which also happen to form an Abelian subalgebra, i.e. copies of $u(1)$. On these grounds and knowing the general boundary conditions, one can expect to obtain the loop algebra of these subalgebras. This construction appears when studying near-horizon boundary conditions for massless higher-spin theory in 3D \cite{Grumiller:2016kcp}. In our construction the boundary conditions will be imposed on the diagonal generators from the construction (\ref{hsconstruction}) in the section (2.1).

For the field $\omega$ as in (\ref{bcsx1}), boundary conditions are given with  $\Omega $ 
\begin{align}
    \Omega(n)=(\mathcal{K}d\phi+\mu dt)L_0+\sum_{s=3}^n(\mathcal{K}_{s}d\phi+\mu_{s}dt)W_{0}^{(s)}. \label{eq41}
\end{align}
$\mathcal{K}$ and $\mathcal{K}_{s}$ in (\ref{eq41}) are arbitrary functions of time and angle and represent the dynamical fields, while $\mu$ and $\mu_{s}$, which also depend on time and angle, represent the chemical potentials. Chemical potentials are assumed not to vary and they are fixed at the boundary.
In comparison to the definition of $\Omega$ in the previous section, here we explicitly include the $d\phi$ and $dt$ components and retain only two generators. In contrast, in the previous section, we commented on the scenario in which we set chemical potentials to zero and included all the generators in the $d\phi$ component. In other words, here we implemented stricter boundary conditions in $d\phi$ component.
 
The infinite set of theories from the CHS theory comes from the class of infinite-dimensional algebras given by the symmetries of higher-order singletons, which are from CFT's of a type $\Box^q\Phi=0$ \cite{Eastwood2008,Joung:2014qya}. If present only the $so(3,2)$-spectrum, this can be written as
\cite{Grigoriev:2019xmp}
\begin{align}
\Box^q\Phi=0:&&\hs\Big|_{so(3,2)}=\bigoplus_{s=0}^{\infty}\bigoplus_{k=0}^{k=q-1} \parbox{90pt}{\RectBRow{9}{7}{$s-1$}{$s-2k-1$}}\, \label{hsalg}
\end{align}
Here, we denote that the decomposition is performed according to the $so(3,2)$ algebra. For each $s$, we obtain an algebra that leads to the corresponding theory. If we remove the $u(1)$ part each such algebra is isomorphic to one copy of a certain $sl(N,\mathbb{R})$ algebra.

Take a look at the example (\ref{hsconstruction}). Each Young diagram in the decomposition belongs to conformal graviton, conformal spin-3 field, and a scalar, respectively. Conformal graviton and conformal spin-3 field are generated by the gauge algebra $sl(5,\mathbb{R})$ that contains 24 generators.  %

Knowing $N$, we can write the boundary conditions as in (\ref{eq41}), with the difference that, in this scenario, we will not have a second copy of $sl(N,\mathbb{R})$ or the corresponding $\overline{\Omega}$. In (\ref{eq41}), the values that $N$ can take are determined by \begin{align} N=\sqrt{\frac{1}{6} \left(2 s_1+3\right) \left(s_1-s_2+1\right) \left(s_1+s_2+2\right) \left(2 s_2+1\right)+1}\label{selectn}.\end{align} Here $s_1$ is the number of boxes in the first row of the Young diagram in (\ref{hsalg}), while $s_2$ is the number of boxes in the second row.
To use the near-horizon bcs in the WL representation, we need to express the WL generators in terms of the PKLD generators. This approach is convenient for comparing the results with those of the $sl(N,\mathbb{R}) \oplus sl(N,\mathbb{R})$ theory. We provide an explicit map from the PKLD basis for the conformal graviton with a vector field, and the conformal graviton with a conformal spin-3 field. The algebra of the conformal graviton with the conformal spin-3 field is isomorphic to the $sl(5,\mathbb{R})$ algebra. For the conformal graviton itself, whose algebra is $so(3,2)$, there is no $N$ for which an isomorphism to any of the $sl(N,\mathbb{R})$ algebras exists. This does not contradict the decomposition (\ref{hsconstruction}) since conformal graviton does not appear by itself from the construction. However, if we want to consider conformal graviton by itself, we can impose the near-horizon boundary conditions to the diagonal generators in representation of the $so(3,2)$ algebra.
Therefore, we impose the near-horizon bcs in this case by choosing diagonal generators directly in the PKLD basis.
 In comparison to most general bcs, fields $\mak_s$ correspond to a certain linear combination of the fields $\tau^{A(k),B(l)}$, while the generators are related via the analogous linear combination which is defined via the map between WL and PKLD basis.\footnote{See  appendices C and D.}  
 
For the theories in (\ref{hsalg}) the asymptotic symmetries and boundary charges are similar to 
 the boundary charges and asymptotic symmetries for the near-horizon boundary conditions of $sl(N,\mathbb{R}) \oplus sl(N,\mathbb{R})$. The difference is that here we can take only one chiral half of  $sl(N,\mathbb{R}) \oplus sl(N,\mathbb{R})$, and we have to select $N$ as described above in (\ref{selectn}). 
The equations of motion for the fields using the near-horizon boundary conditions (\ref{eq41}) read 
\begin{align}
    \partial_t\mathcal{K}=\partial_{\phi}\mu, &&   \partial_t\mak_{0}^{(3)}=\partial_{\phi}\mu_3, && ... && \partial_t\mathcal{K}_{0}^{(N)}=\partial_{\phi}\mu_{N}.\label{eomsls}
\end{align}
The form (\ref{eq41}) is maintained by the asymptotic symmetries defined by parameters valued in a Lie algebra. The gauge parameters take the form 
\begin{align}
\epsilon=\xi L_0+ \sum_{s=3}^N\xi_{s}W_0^{(s)}\label{eq43} 
\end{align}
for $\partial_t\xi=\partial_t\xi_3=...=\partial_t\xi_{N}=0$, if the fields transform as 
\begin{align}
    \delta\mathcal{K}=\partial_{\phi}\xi, && \delta\mak_{0}^{(3)}=\partial_{\phi}\xi_3, && ... && \delta \mathcal{K}_{0}^{(N)}=\partial_{\phi}\xi_{N}.\label{trafosls}
\end{align}
We can write the canonical generators of the asymptotic symmetries as 
\begin{align}
    \mathcal{Q}[\eta,..,\eta_{(n)},N]=-\frac{k}{4\pi}\int d\phi\left(\xi(\phi)\mathcal{K}(\phi)+\sum_{s=3}^N\alpha_s\xi_{s}(\phi)\mathcal{K}_{0}^{(s)}(\phi))\right)\label{eq46}
\end{align}
for 
\begin{align}
\alpha^{(N)}_s=\frac{48(s-1)!^4}{(2s-1)!(2s-2)!}\prod_{i=2}^{s-1}(N^2-i^2),\label{alpha}
\end{align}
where $\alpha_s^N$ are coming from the invariant bilinear form of the algebra which can also be seen from the definition of the charges (3.4).
Calculating Poisson brackets of the charges $\delta_YQ[X]=\{Q[X],Q[Y]\}$ and the variation of the fields, 
we obtain the algebra
\besubeqs\label{eq276}
\begin{align}
    \{\mathcal{K}(\phi),\mathcal{K}(\phi')\}&=-\frac{4\pi}{k}\delta'(\phi-\phi')\\
...  \\
    \{\mathcal{K}_{0}^{(N)}(\phi),\mathcal{K}_{0}^{(N)}(\phi')\}&=-\frac{4\pi}{\alpha_Nk}\delta'(\phi-\phi').
    \end{align}
\esubeqs
If we expand the charges in the Fourier modes using 
\begin{align}
\mathcal{K(\phi)}=\frac{2}{k}\sum_{p=-\infty}^{\infty}J_pe^{ip\phi}, && \mathcal{K}_{0}^{(3)}(\phi)=\frac{2}{\alpha_4k}\sum_{p=-\infty}^{\infty}J^{(3)}_pe^{ip\phi}, &&... && \mathcal{K}_{0}^{(N)}(\phi)=\frac{2}{\alpha_Nk}\sum_{p=-\infty}^{\infty}J^{(N)}_pe^{ip\phi},\label{expsls}
\end{align}
the algebra (\ref{eq276}) gives ASA
\besubeqs
\begin{align}
i\{J_p,J_m\}&=\frac{k}{2}p\delta_{m+p,0}  \\
 i\{J^{(3)}_p,J^{(3)}_m\}&=\frac{\alpha_3k}{2}p\delta_{m+p,0} \\
 ... \nonumber\\
 i\{J^{(N)}_p,J^{(N)}_m\}&=\frac{\alpha_Nk}{2}p\delta_{m+p,0}.
\end{align}\label{asaslg}
\esubeqs
For arbitrary $N$, this algebra consists of $u(1)$ currents with levels $\frac{1}{2}k, \ldots, \frac{1}{2}\alpha_N k$, which corresponds to expected ASA, or loop algebra of Abelian subalgebra formed by the generators kept in boundary conditions. Brackets that are zero are not shown.

To determine the asymptotic symmetries for a theory with a general conformal spin-$s$, one must find a general map between the PKLD and WL bases. The maps between PKLD and WL basis are determined on a case-by-case basis, which makes it not straightforward to generalize to the map for the $sl(N)$ case. 
Below, we explicitly present the asymptotic symmetries for the first few theories, for the conformal graviton, the conformal spin-1 field with the graviton, and the conformal spin-3 field with the graviton.
Each of these cases is written using the specific map. This means that they should not be compared to each other, however the conclusions should be made within each case.
I.e., in the case of a theory with conformal spin-1 field and a graviton, we can consider how the entropy changes switching off spin-1 field; and in the case of conformal spin-3 field and a graviton we can consider the same by switching off spin-3 field. However, the theories should not be compared directly, only conceptually, because of the different  embeddings.


\subsection{so(3,2): conformal spin-2 field}

If we apply the near-horizon boundary conditions on conformal gravity in 3D, we need to identify the diagonal generators in the PKLD basis. To express the algebra of $so(3,2)$
\besubeqs\label{LPDK}
\begin{align}
[\DDD,\PPP^a] &= -\PPP^a,, & [\LLL^{ab}, \PPP^c] &= \PPP^a \eta^{bc} - \PPP^b \eta^{ac}, \\
[\DDD, \KKK^a] &= \KKK^a,, & [\LLL^{ab}, \KKK^c] &= \KKK^a \eta^{bc} - \KKK^b \eta^{ac}, \\
[\PPP^a, \KKK^b] &= -\LLL^{ab} + \eta^{ab} \DDD,, & [\LLL^{ab}, \LLL^{cd}] &= \LLL^{ad} \eta^{bc} + \text{three more}, \label{cgalg}
\end{align}
\esubeqs 
 we use the standard representation in terms of $\gamma$ matrices. For the generators $\PPP_i,\KKK_i,\LLL_{ij},\DDD$ explicitly, we use the representation in the Appendix D, in the equation  (\ref{matex1}). %

To impose the near-horizon boundary conditions, we note that two diagonal generators exist and can be used. One of them is the $\DDD$ generator, while the other is a generator of Lorentz rotations, which we denote as $\LLL_d$ to indicate that it is represented by a diagonal matrix.
For $\Omega$ we can write the ansatz 
\begin{align}
    \Omega=(\mathcal{K}(\phi,t) d\phi+\mu(\phi,t) dt)\DDD+(\mathcal{K}_d(\phi,t) d\phi+\mu_d(\phi,t) dt)\LLL_d \label{omso32}.
\end{align}
Since the exact form of the group element $b$ is not relevant for the first part of the analysis in the Chern-Simons form, where $\delta b = 0$, we assume it to be a function of $\rho$, i.e., $b = b(\rho)$.
Here, we define the charges as $\mathcal{K}$ and $\mathcal{K}_d$, which are allowed to vary:
\begin{align}
\delta\Omega_{\phi} &= \delta\mathcal{K}(\phi,t)\DDD + \delta\mathcal{K}_d(\phi,t)\LLL_d, &&
\delta\Omega_t = 0,
\end{align}
while the variations of the chemical potentials $\mu$ and $\mu_d$ vanish.

The equations of motion lead to expressions similar to those in (\ref{eomsls}), which are familiar for the chosen gauge and boundary conditions:
\begin{align}
\partial_t \mathcal{K} = \partial_{\phi} \mu, && \partial_t \mathcal{K}_{d} = \partial_{\phi} \mu_d.
\end{align}
 The parameter $\epsilon$, which is valued in the Lie algebra and keeps the asymptotic symmetries takes the form 
 \begin{align}
     \epsilon=\xi \DDD+\xi_{d}\LLL_d
 \end{align}
 where $\partial_t\xi=\partial_t\xi_d=0$. The fields $\mathcal{K}$ and $\mathcal{K}_d$ then transform as 
 \begin{align}
 \delta\mathcal{K}=\partial_{\phi}\xi,  &&      \delta\mathcal{K}_d=\partial_{\phi}\xi_d.
 \end{align}
  Using the variation of the canonical boundary charges (\ref{varcharge}) one can obtain 
  \begin{align}
      \delta Q[\epsilon]=\frac{k}{2\pi}\oint d\phi(\xi\delta\mathcal{K}\langle \DDD\DDD\rangle+\xi_d\delta\mathcal{K}_d\langle \LLL_d\LLL_d\rangle)
  \end{align}
  and the conserved charges that are associated with the asymptotic symmetries are
  \begin{align}
       Q[\xi,\xi_d]&=\frac{k}{2\pi}\oint d\phi(\xi(\phi)\mathcal{K}(\phi)\langle \DDD\DDD\rangle+\xi_d(\phi)\mathcal{K}_d(\phi)\langle \LLL_d\LLL_d\rangle)\nonumber\\
       &=\frac{k}{2\pi}\oint d\phi(2\xi(\phi)\mathcal{K}(\phi)+\frac{3}{2}\xi_d(\phi)\mathcal{K}_d(\phi)).
  \end{align}
Using the Poisson brackets as above and using the expansion of the charges in the Fourier modes we obtain the ASA  \begin{align}
i\{J_n,J_m\}&=k n\delta_{m+n,0} && i\{J^{(d)}_n,J^{(d)}_m\}=\frac{3k}{4}n\delta_{m+n,0}.
\end{align}
which corresponds to $u(1)$ currents, while $J_n$ and $J_n^{(d)}$  are Fourier modes corresponding to charges $\mak$ and $\mak_d$, respectively.

For the spherically symmetric solution and certain set of assumptions satisfied by the near-horizon bcs, the entropy \cite{Perez:2012cf,Perez:2013xi, deBoer:2013gz} is given by
\begin{align}
    S=-\frac{k_N}{\pi}Im\beta\int d\phi \langle\Omega_{\phi}\Omega_t\rangle.\label{entropygen}
\end{align}
To evaluate this, we have to use the definition for $\Omega$ (\ref{omso32}) combined with the definition $\Omega=\Omega_{\phi}d\phi +\Omega_t dt$, and the definition for trace of $\langle \DDD \DDD\rangle$ and $\langle \LLL_d \LLL_d\rangle$ matrices. This is where the dependence on the embedding or the map between PKLD and WL bases appears in the relative entropy, assuming that the entropy of the lowest-spin particle is normalized.
The entropy for these bcs is given by
\begin{align}
    S=-2\frac{k_N}{\pi} Im\left[ \beta\int d\phi(\mu\mathcal{K}-\mu_d\mathcal{K}_d)\right].
\end{align}
The chemical potentials can be evaluated as it was done in \cite{Grumiller:2016kcp}. Here we define 
the contractible circle $\tau=it$ and Hawking temperature $T=1/\beta$. 
For the near-horizon bcs one can find the chemical potentials from the holonomy around the contractible circle \cite{Bunster:2014mua} 
\begin{align}
e^{\int \Omega_t}=-\mathds{1}
\end{align}
For the $\Omega_t$ from (\ref{omso32}) one obtains a solution
\begin{align}
\mu=-\frac{2}{\beta}(m+n+1)\pi, && \mu_d=\frac{2}{\beta}(m-n)\pi
\end{align}
for $n,m\in\mathbb{Z}$.  Which from $S=-k_N\beta(\mu\mathcal{K}-\mu_d\mathcal{K}_d)$ leads to \begin{align} S=-2k_N \left[ -(m+n+1)\pi\mathcal{K}-(m-n)\pi\mathcal{K}_d \right] \label{entropyso32}\end{align}
We can notice that when the chemical potential for the dilatation is zero, it means $m=-n-1$, in which case the entropy becomes $S=-2k_N(2n+1)\pi\mathcal{K}_d$. Since this corresponds to the generator that is within Poincar\'e algebra, for $n=0$, one can think of it as a charge that corresponds to $L_0$ generator of one half of the $sl(2)\oplus sl(2)$ of Einstein gravity in three dimensions. Or, depending on the embedding of the $sl(2)\oplus sl(2)$, it can be a charge corresponding to linear combination of $L_0^+$ and $L_0^-$ generators (for $x^+=t+\phi$ and $x^-=t-\phi$.). The branch with entropy corresponding to the charges of the generators $L_0^+$ and $L_0^-$ in Einstein gravity, is the branch of higher-spin black flowers that is  continuously connected to BTZ black hole. 

{\bf Metric formulation--(simplest) example 1.} To transition to the metric formulation we follow the prescription for the near-horizon bcs.  We write (\ref{omso32}) in  (\ref{bcsx1}). This leads to $\omega$ defined by
\begin{align}
\omega=b^{-1}[b_1\DDD+b_2\LLL_d]b+b^{-1}\partial_{\rho}bd\rho 
\end{align}
for $b_1=\mu dt+\mathcal{K}d\phi$, $b_2=\mu_d dt+\mathcal{K}_d d\phi$. Here, $\mu,\mu_d, \mathcal{K},\mathcal{K}_d$ are functions of ($t,\phi$).

We choose $b$ to be a function of $\rho$, which partially fixes the gauge. It must also belong to a group that we are considering, in this case, $SO(3,2)$. A common choice is 
 $b(\rho)=e^{Y(\rho)}$. We refer to $b$ as a projector because it projects the solution from the first-order formulation to the metric formulation.
 We define $Y(\rho)$ as 
 \begin{align}Y(\rho)=\rho \PPP_y+\PPP_x + \tilde{c}_1 \DDD.\label{ex1}\end{align} 
 (The indices $i,j$ of the generators $\PPP_i,\KKK_i,\LLL_{ij}$ we shift with the metric $\delta_{ij} = \mathrm{diag}(1, 1, -1)$, and generators are represented by matrices, as described in Appendix D.)
The definition of the metric in conformal gravity then follows from the $g_{\mu\nu} = \langle e_{\mu}e_{\nu}\rangle$, where $e_{\mu}{}^a$ is read out from fields associated with the generator of translations. The metric is of course invariant under Weyl rescaling, that we in reading out the metric keep to be one. However in bringing the metric to more familiar form we keep this in mind.
 For the group element $b$ defined by defined by $Y(\rho)$ in (\ref{ex1})
we obtain the dreibein 
\begin{align}
 e^a_{\mu} =  c_1 \left(
\begin{array}{ccc}
 0 & 1 & 0 \\
   -\mak & -\mathcal{K} \rho  & \mathcal{K}_d \rho \\
   -\mu & -\mu  \rho  &  \mu_d\rho  \\
\end{array}
\right)
\end{align}
which after identification defines a metric. To more easily transform a metric to a desired form we assume constant fields and chemical potentials. Bringing the metric to ADM form it leads to BTZ-like metric, with $g_{rt}=g_{r\phi}=0$,  $g_{\phi\phi}=r^2$ and
\besubeqs
\begin{align}
g_{rr}&=\frac{r^2 }{\left(\mak^2-r^2\right) \left(\mak_d^2-r^2\right)}\label{ex1grr}\\
g_{\phi t}&=\frac{ f_2'(t) \left(\mak \mu  \left(r^2-\mak_d^2\right)+\mak_d \mu _3 \left(\mak^2-r^2\right)\right)}{\mak_d^2-\mak^2}+r^2 f_1'(t)\\
g_{tt}&=\frac{2 f_2'(t)\bigl( f_1'( t) \left(\mak \mu  \left(r^2-\mak_d^2\right)+\mak_d \mu _3 \left(\mak^2-r^2\right)\right)- f_2'( t) \left(\mu ^2 (\mak_d^2-r^2) -\mu _3^2 \left(\mak^2-r^2\right)\right)\bigr)}{{\mak^2-\mak_d^2}}\nonumber\\&+ r^2 f_1'( t){}^2 \end{align}\label{simpex}\esubeqs
Here $c_1=\frac{-1+e^{\tilde{c}_1}}{\tilde{c}_1}$ and $f_1$ and $f_2$ are functions of $t$ such that $f_i(t)=\tilde{f_i}((-1+e)t)$ for $i=1,2$, and we rescaled chemical potentials with $(e-1)$. The curvature $R=-\frac{6}{c_1^2}$  for the solution (\ref{simpex}) points to an anti-de Sitter solution.
Fixing the functions $f_1$ to $ t$ and $f_2$ to $t/(-1+e)$ the components $g_{\phi t}$ and $g_{tt}$ become 
\besubeqs
\begin{align}
g_{\phi t}&=\frac{ \left(\mak \mu  \left(r^2-\mak_d^2\right)+\mak_d \mu _3 \left(\mak^2-r^2\right)\right)}{\mak_d^2-\mak^2}+r^2 \\
g_{tt}&=\frac{2 \bigl(  \left(\mak \mu  \left(r^2-\mak_d^2\right)+\mak_d \mu _3 \left(\mak^2-r^2\right)\right)- \left(\mu ^2 (\mak_d^2-r^2) -\mu _3^2 \left(\mak^2-r^2\right)\right)\bigr)}{{\mak^2-\mak_d^2}}+r^2 \label{e434}
\end{align}
\esubeqs
Switching off both chemical potentials would lead to degenerate metric with zero determinant.
From $g_{rr}$ one can see that solution has two real eigenvalues which would indicate that it belongs to the particular class of the metric which is $I_b$ according to the classification described in \cite{Lovrekovic:2025dwd}. 

To compare the solution with the case where entropy is proportional to $\mak_d$ when, $\mu=0$, we take $\mu$ to be zero in the solution above. 
Fixing the chemical potential $\mu_d$ to $\mu_d=\frac{\mak^2-\mak^2_d}{\mak_d}$ and $t\rightarrow \frac{\mak_d}{\mak }t$ we obtain the AdS solution exactly of the BTZ form. We have one horizon which is $r_{1}=\mathcal{K}_d$ while another is $r_2=\mathcal{K}$. Setting $r_+=r_1$ and $r_-=r_2$ in the standard BTZ metric \cite{Banados:1992gq} depicts the solution.

{\bf Metric formulation -- example 2.} In this example we examine the influence of $\rho$ dependency in the projector $b$.
We define the $Y(\rho)$ as $Y(\rho)=a(\rho)\DDD+b(\rho)\PPP_y+c(\rho)\PPP_x$. The dreibein becomes
\begin{footnotesize}
\begin{align}
   e_{\mu}{}^{a}=\frac{(-1+e^{a(\rho)})}{a(\rho)}\left(
\begin{array}{ccc}
 c'(\rho )+c(\rho ) a'(\rho ) \left(\frac{1}{e^{a(\rho )}-1}-\frac{1}{a(\rho )}\right) & b'(\rho )+b(\rho )a'(\rho ) \left(\frac{1}{e^{a(\rho )}-1}-\frac{1}{a(\rho )}\right)  & 0 \\
  -\mak c(\rho )   & -\mak b(\rho ) &  \mak_d b(\rho ) \\
  -\mu  c(\rho )    & -\mu  b(\rho ) &  \mu _d b(\rho )  \\
\end{array}\right)
 \end{align}
\end{footnotesize}
and gives the metric $g_{ij}$ with the following non-vanishing terms 
\besubeqs \label{gengen}
\begin{align}
g_{\rho\rho}=&\frac{1}{{a(\rho )^2 c_1(\rho )}}\left[\left(b(\rho ) \left(c_1(\rho )-1\right) a'(\rho )-a(\rho ) c_1(\rho ) b'(\rho )\right){}^2 \nonumber \right. \\& \left.+\left(c(\rho ) \left(c_1(\rho )-1\right) a'(\rho )-a(\rho ) c_1(\rho ) c'(\rho )\right){}^2\right]\\
g_{\rho\phi}=&-\frac{c_1(\rho)\mak}{a(\rho)}\left[ b(\rho)\left(a(\rho ) c_1(\rho ) b'(\rho )-b(\rho )\left(c_1(\rho )-1\right) a'(\rho )\right) \right. \nonumber \\ & \left. +c(\rho ) \left(a(\rho ) c_1(\rho ) c'(\rho )-c(\rho ) \left(c_1(\rho )-1\right) a'(\rho )\right)\right]\\
g_{\rho t}=&-\frac{c_1(\rho)\mu}{a(\rho)}\left[ b(\rho)\left(a(\rho ) c_1(\rho ) b'(\rho )-b(\rho )\left(c_1(\rho )-1\right) a'(\rho )\right) \right. \nonumber \\ & \left. +c(\rho ) \left(a(\rho ) c_1(\rho ) c'(\rho )-c(\rho ) \left(c_1(\rho )-1\right) a'(\rho )\right)\right]\\
g_{\phi\phi}=&c_1(\rho )^2 \left(b(\rho )^2 (\mak^2-\mak_d^2) +\mak^2 c(\rho )^2\right)\\
g_{\phi t}=& c_1(\rho )^2 \left(b(\rho )^2 (\mak \mu-\mak_d\mu_d )+\mak \mu c(\rho )^2\right)\\
g_{t t}=& c_1(\rho )^2 \left(b(\rho )^2 (\mu^2-\mu^2_d) +\mu^2 c(\rho )^2\right).
\end{align}
\esubeqs
for $c_1(\rho)=\frac{e^{a(\rho )}-1}{a(\rho)}$ arbitrary parameter. The above metric contains the fields and chemical potentials with their initial dependency on the coordinates.
Setting them to constant values lets us find
suitable choices of the $c(\rho)$ and $b(\rho)$ that reduce the metric to one of the simpler forms. Here, we determine the function $b(\rho)$ by requiring $\rho^2d\phi^2$. We then transform the coordinates $\phi\rightarrow \phi+ \frac{ \mu_d\left[2  (a(\rho )+\log (a(\rho )))- \log \left( a(\rho )^2(\rho^2-\mak_d^2 c_1(\rho)^2 c(\rho )^2)\right)\right]}{2 \mak_d \mu-2 k \mu_d}+c_2(T)$ and $t\rightarrow \frac{ -\mak_d\left[2  (a(\rho )+\log (a(\rho )))- \log \left( a(\rho )^2(\rho^2-\mak_d^2 c_1(\rho)^2 c(\rho )^2)\right)\right]}{2 \mak_d \mu-2 k \mu_d}+c_1(T)$. We further rescale the metric with the numerator in $g_{\rho\rho}/\rho^2$ and once more, fix $g_{\phi\phi}=\rho^2$, this time using the function $c(\rho)$. We simplify the result by setting $a(\rho)=1$, which leads to metric  
\begin{align}
  ds^2=f_{\rho\rho}(\rho) d\rho^2+ f_{tt}(\rho) dt^2 -2 \rho^2   n(\rho)dt d\phi+\rho^2 d\phi^2\label{gbtz}
\end{align}
\besubeqs \label{genspec}
for \begin{align}
    f_{tt}(\rho)&=\frac{1}{\mak^2-\mak^2_d}\left[(\mak^2-\mak_d^2) \left(\mak^2 \mu_d^2-\mak_d^2 \mu^2+\rho ^2 (\mu^2-\mu_d^2)\right) \right. \\ & \left. -2 q_2 (\mak_d \mu-\mak \mu_d) \left(-\mak^2 \mak_d \mu_d+\mak \mak_d \rho  q_1 \left(\rho  q_1-2\right) (\mak_d \mu-\mak \mu_d)+\mak \mu (\mak_d^2-\rho^2 ) +\mak_d \rho ^2 \mu_d\right)\right.\nonumber \\& \left. +\rho ^2 q_2^2 (\mak_d \mu-\mak \mu_d)^2+\rho  q_1 (\mak^2-\mak_d^2)  \left(\rho  q_1-2\right) (\mak^2 \mu^2_d-\mak^2_d \mu^2)  \right]\nonumber\\
    f_{\rho\rho}(\rho)&=\frac{\rho ^2 }{\left(\mak \rho  q_1-\mak-\rho \right) \left(\mak \rho  q_1-\mak+\rho \right) \left(\mak_d \rho  q_1-\mak_d-\rho \right) \left(\mak_d \rho  q_1-\mak_d+\rho \right)} \label{horiz}\\
    n(\rho)&=\frac{\rho  \left(\mak \mak_d q_1 \left(\rho  q_1-2\right)-\rho  q_2\right) (\mak \mu_d-\mak_d \mu)+\rho ^2 (\mak \mu-\mak_d \mu_d)+\mak \mak_d (\mak \mu_d-\mak_d \mu)}{\rho ^2 (\mak_d \mu-\mak \mu_d)}
\end{align}
\esubeqs
while the $f_{tt}(\rho)=\rho^2n(\rho)^2-1/f_{\rho\rho}(\rho)$, and $q_1$ and $q_2$ are free parameters. 
The parameter $q_2$ we obtained by fixing the function $c_2(T)=q_2 T$, while we fixed the function $c_1$ to  $c_1(T)=\frac{\mak_d^2-\mak^2}{\mak \mu_d-\mak_d \mu} T $.
The solution satisfies the Cotton tensor, however its Ricci scalar is not constant,\\ $R=-6+\frac{2 q_1}{\rho ^3} \left[\left(-2 \mak^2 \left(\mak_d^2+\rho ^2\right) +\rho  q_1 \left(3 \rho ^2 \left(\mak^2+\mak_d^2\right)+\mak^2 \mak_d^2 \rho  q_1 \left(4-3 \rho  q_1\right)+\mak^2 \mak_d^2\right)-2 \mak_d^2 \rho ^2\right)\right]$. Taking $\rho\to \infty$ the Ricci scalar goes to value determined by the fields and an arbitrary constant, $R=-6 \left(\mak^2 q_1^2-1\right) \left(\mak_d^2 q_1^2-1\right)$.
The metric written in this form has two real eigenvalues for general values of $\mak$, $\mak_d$ and $q_1$. For the special case when $\mak,\mak_d,q_1>0$, $\mak q_1>1$  and $\mak_d q_1>1$  we have four positive real radii. These are visible from (\ref{horiz}).
If we fix the $b(\rho)$ in (\ref{gengen}) to get $\rho^2$ for the $g_{\phi\phi}$ term in the metric, transform the coordinates such that $g_{\rho\phi}$ and $g_{\rho t}$ vanish, we can fix  $c(\rho)=1/(-1 + e^a(\rho))$ and $a(\rho)=1+\frac{c_1}{\rho }$ to get the similar form of the metric. Suitable choice of   parameters in such metric can also be chosen to give four positive radii. The metric gives the Cotton tensor equal to zero.
It would be interesting to further analyze this metric.

 If in the above metric, we set $\mak=0$, $q_2=\frac{\mu_2}{\mu}$ and rescale a metric with some constant $a_1$, after absorption of $a_1$ into $\phi$ and $a_1^2$ into $t$ coordinates, we recover the three-dimensional analog \cite{Oliva:2009hz} of the four dimensional MKR (Mannheim-Kazanas-Riegert) solution \cite{Mannheim:1988dj,Riegert:1984zz}. The metric is then $ds^2=d\rho^2/f(\rho)-f(\rho)dt^2+\rho^2d\phi^2$, for $f(\rho)=a_1^2\mak_d^2-2a_1^2\mak_d^2q_1 \rho+a_1^2(\mak_d^2q_1^2-1)\rho^2$. Depending on the choice of $\mak_d$, $q_1$ and $a_1$, the solution can contain general form of eigenvalue with real and imaginary part, contained in the class $I_a$. We also elaborate on this metric in the Appendix B.

From the general form of the metric we can also obtain de Sitter solution with one positive horizon, which belongs to the $I_d$ class in classification \cite{Lovrekovic:2025dwd}. Similar kind of metrics have been analyzed in a different context in \cite{Emparan:2022ijy}.  They considered backreaction of quantum fields as perturbation in conical $dS_3$, which allowed a formulation of a black hole in the $dS_3$ spacetime.

In the Appendix C we show how changing dependence on $\rho$ in front of a generator in the projector $b$ influences the metric, and which boundary conditions lead to Lobachevksy solution.

\subsection{$sl(4)$: conformal spin-2 and spin-1 field}

The algebra of conformal graviton and a vector field in the PKLD basis is isomorphic to $sl(4)$ algebra. The commutation relations in PKLD basis consist of the commutation relations (\ref{pkldv}). Explicitly after decomposition in the light-cone coordinates, these commutation relations become (\ref{LPDK}) and 
 \begin{align}
    [\PPP_a,\ttt_+]&=0\,, & [\PPP_a,\ttt_-]&=\ttt_a\,,& [\PPP_a,\ttt_c]&=-\eta_{ac}\ttt_+ \,, \nonumber\\ 
    [\LLL_{ab},\ttt_+]&=0\,, & [\LLL_{ab},\ttt_c]&=\eta_{bc}\ttt_a-\eta_{ac}\ttt_b\,,& [\LLL_{ab},\ttt_-]&=0 \,, \nonumber\\
    [\KKK_{a},\ttt_+]&=\ttt_a \,,& [\KKK_{a},\ttt_b]&=-\eta_{ab}\ttt_- \,, &[\KKK_{a},\ttt_-]&=0 \nonumber\\
    [\DDD,\ttt_+]&=-\ttt_+ \,, &[\DDD,\ttt_a]&=0 \,, & [\DDD,\ttt_-]&=\ttt_- \nonumber\\
    [\ttt_+,\ttt_-]&=-\DDD \,, & [\ttt_a,\ttt_+]&=\PPP_{a} \,, & [\ttt_a,\ttt_-]&=\KKK_{a},\nonumber\\ 
    [\ttt_a,\ttt_b]&=\LLL_{ab}. &&&& \label{commute1}
    \end{align}
The diagonal matrices in the WL basis are given by linear combination of matrices in PKLD basis. 
The PKLD basis for the $sl(4)$ group is described in Appendix D. There, we present the $sl(4)$ group in both the WL and PKLD representations  and provide a map from the PKLD to the WL representation.
To study near-horizon bcs and be able to compare this with the earlier work it is more convenient to consider the matrices in WL basis. In WL basis,
$sl(4)$ algebra closes the commutation relations 
\begin{align}
    [\LLL_n,\LLL_m]=(n-m)\LLL_{n+m} && [\LLL_n,\VVV_{m}^{(s)}]=((s-1)n-m)\VVV_{n+m}^{(s)}
\end{align}
where the indices in $\VVV^{(4)}_m$ and $\VVV^{(3)}_n$ are
 $m=0,\pm1,\pm2,\pm3$ and $n=0,\pm1,\pm2$ respectively. The generators $\LLL_n$ are generators of $sl_2$ algebra.

Once we know the map between the generators in PKLD basis and the WL basis (Appendix D), we can define the near-horizon bcs by specifying $n=4$ in the general case (\ref{eq41})
\begin{align}
\Omega(n)=(\mak d\phi+\mu dt)\LLL_0+(\mak_{0}^{(3)}d\phi+\mu_3dt)\VVV_0^{(3)}+(\mak_{0}^{(4)}d\phi+\mu_4dt)\VVV_0^{(4)}.\label{eq441}
\end{align}
 The $\epsilon$ is defined from (\ref{eq43})
\begin{align}
\epsilon=\xi \LLL_0+\xi_3\VVV_0^{(3)}+\xi_4\VVV_0^{(4)}.
\end{align}
 By defining $n=4$ in (\ref{eq46}) we can also read out the charge, which is 
\begin{align}
    Q[\xi,\xi_3,\xi_4]=-\frac{k}{4\pi}\int d\phi\left(\xi(\phi)\mak(\phi)+\alpha_3\xi_3(\phi)\mak_{0}^{(3)}(\phi)+\alpha_4\xi_4(\phi)\mak_{0}^{(4)}(\phi) \right).
\end{align}
ASA, is the algebra (\ref{asaslg}) for charges $\mak,\mak_{0}^{(3)}$ and $\mak_{0}^{(4)}$, and corresponding $u(1)$ currents the algebra consists of, have   levels $\frac{1}{2}k$, $\frac{1}{2}\alpha_3k$ and $\frac{1}{2}\alpha_4 k$.

To evaluate the entropy of this solution, we use our definition for $\Omega$ (\ref{eq441}) as above, and definitions for traces of $\langle \LLL_0 \LLL_0\rangle$ and $\langle \VVV_0^{(s)}\VVV_0^{(s)}\rangle$ matrices, and formula (\ref{entropygen}).
This leads to 
\begin{align}
    S&=-\frac{k_N}{\pi}Im\left[\beta\int d\phi \left[5\mu\mathcal{K}+16\mu^{(3)}\mathcal{K}_0^{(3)}+\frac{36}{5}\mu^{(4)}\mathcal{K}_0^{(4)}\right]\right].\nonumber\\
&=-2k_N\beta\left[5\mu\mathcal{K}+16\mu^{(3)}\mathcal{K}_0^{(3)}+\frac{36}{5}\mu^{(4)}\mathcal{K}_0^{(4)}\right].
\end{align}
Here we have fields of spin 1 and spin 2. We can evaluate the chemical potentials in the similar way as for $so(3,2)$ 
\begin{align}
\mu=\frac{2 \pi  (l-m-2 n-1)}{5 \beta } && \mu_3=\frac{\pi  (l+n+1)}{2 \beta }, && \mu_4=\frac{\pi  (2 l+3 m+n+3)}{3 \beta }.
\end{align}
Which gives for the entropy
\begin{align}
S=&-\pi k_N\frac{4  }{5 \beta }  \left(5 \mak  (l-m-2 n-  1) \right. \nonumber\\& \left.+6 \mak_0^{(4)} (2 l+ 3 m+n+3)+20 \mak_0^{(3)} (l+n+1 )\right). \label{entropysl4}
\end{align} 
We know from the map between the PKLD and WL bases (\ref{mapsl4}) that $\mak$ is the charge for the generator that is a linear combination of dilatations and one of the conformal vector generators. Since dilatations are not part of the Poincar\'e group, we set $\mu$ to zero. The same is true for $\mak_0^{(4)}$, so we set $\mu_4$ to zero as well. This leads to the relations $m = -1 - l$ and $n = l$.
In this case, the entropy that arises from one of the generators of the Poincar\'e algebra is described by the charge $\mak_0^{(3)}$. The corresponding generator is proportional to $\LLL_d$ (from the conformal graviton case). Specifically, for $\mu = \mu_4 = 0$, the entropy is $S = -16\pi k_N (1 + 2l) \mak_0^{(3)}$. This entropy, up to a proportionality factor, agrees with the solution in the $so(3,2)$ case.

\textbf{Metric formulation.} 
To consider metric formulation,
similarly as in the previous section, we have to 
identify the dreibein $e_{\mu}^a$ from $\omega$, while we obtain $\omega$ by inserting (\ref{eq441}) in (\ref{bcsx1}).
If we denote $b_1=\mathcal{K}d\phi+\mu dt,b_2=\mathcal{K}^{(3)}_0d\phi+\mu_3 dt,b_3=\mathcal{K}^{(4)}_0d\phi+\mu_4 dt$, in our decomposition we can write 
\begin{align}
\omega=b^{-1}[b_1\LLL_0+b_2\VVV_0^{(3)}+b_3\VVV_0^{(4)}]b+b^{-1}\partial_{\rho}bd\rho \label{e274}.
\end{align}
 If we use  group element $b$ built from $ Y=\rho \PPP_y+\PPP_x+\tilde{c}_1\DDD  $ 
we obtain the dreibein
\begin{align}
    e_{\mu a}=c_1 \left(
\begin{array}{ccc}
 0 & 1 & 0 \\
\frac{-2}{5}  \left(5 \mathcal{K}-3 \mak_0^{(4)}\right)   & \frac{-2}{5}  \left(5 \mathcal{K}-3 \mak_0^{(4)}\right) \rho  & 
 -4 \mak_{0}^{(3)} \rho    \\
\frac{-2}{5} \left(5 \mu -3 \mu _4\right)    & \frac{-2}{5}  \left(5 \mu -3 \mu _4\right) \rho  &    -4 \mu _3 \rho\\
\end{array}
\right)
\end{align} where $c_1\equiv\frac{(-1+e^{\tilde{c}_1})}{\tilde{c}_1}$. The dreibein is of the same form as the dreibein in the example 1 from the previous subsection about conformal spin-2 field, with the identification \begin{align}\mathcal{K}(so(3,2))=-\frac{2}{5}[5\mathcal{K}(sl(4))-3\mak_0^{(4)}(sl(4))], &&
 \mathcal{K}_d(so(3,2))=-4\mak_{0}^{(3)}(sl(4)), \\ \mu(so(3,2))=-\frac{2}{5}[5\mu(sl(4))-3\mu(sl(4))],  && \mu_d(so(3,2))=-4\mu_3(sl(4)). \end{align}
The procedure to obtain the dreibein $e_{\mu}^a$ from the CS connection valued in the  $so(3,2)$ and $sl(4)$ (or $sl(n)$) algebras is analogous.
The difference in $e_{\mu}^a$ arises due to additional terms in the boundary conditions (\ref{eq441}).
 Also, new additional contribution to the metric will appear for example, when group element $b$ is formed using $\ttt_{+}$ generator.

Using $g_{\mu\nu} =  e^{a}_{\mu} e_{\nu}^b\eta_{ab}$ leads to a metric with a Ricci scalar of $-\frac{6}{c_1^2}$.
After transforming the coordinates into the ADM form and setting $\rho = r$, the metric reads exactly as in (\ref{simpex}), with the above defined changes in fields and chemical potentials.

When we set the functions $f_1(t)\rightarrow t$ and $f_2(t)\rightarrow t$ (analogous to those in (\ref{simpex})) and the chemical potentials $\mu=\mu_4=0$ and $\mu_3=\frac{(\mak-\mak_{0}^{(4)})^2-(\mak_{0}^{(3)}){}^2}{\mak_{0}^{(3)}}$ we obtain the metric 
\begin{align}
ds^2&=   \frac{c_1^2 r ^2 dr ^2}{((\mak_{0}^{(3)}){}^2-r^2 )  \left((\mak_0^{(4)}-\mathcal{K})^2-r ^2\right)}+dt^2 \left((\mak_{0}^{(3)}){}^2+(\mak_0^{(4)}-\mak)^2-r ^2\right)+\nonumber\\&+2 \mak_{0}^{(3)}   (-\mak_0^{(4)}+\mathcal{K})dt d\phi+r ^2 d\phi ^2\label{sl4met}
\end{align}
 that has the form of the BTZ metric. One can notice that the existence of the second horizon depends on the behaviour of the $\mak_0^{(4)}$ and $\mathcal{K}$. The chemical potential $\mu_3$ got absorbed during the redefinition of the coordinates, and brought to the fact that resulting metric depends on the fields $\mathcal{K}$ and $\mathcal{K}_0^{(i)}$, $i=3,4$, and constant $c_1$.

\subsection{$sl(5)$: conformal spin-2 and spin-3 field}
Similar to the two previous theories, we first find a map from our generators in the PKLD basis to the WL basis. The $sl(5)$ algebra in the WL basis corresponds to the algebra of the conformal graviton and a spin-3 field in PKLD basis.

From the definitions (\ref{alg2}) and (\ref{eq15}), we can determine the 5x5
 matrices for the conformal graviton and the spin-3 field, which can be mapped to a WL representation. 
The matrices satisfy the condition $2\SSDD=-(\SSLL_{xx}+\SSLL_{yy}+\SSLL_{tt})$, and the commutation relations 
\besubeqs \label{com1e3}
\begin{align}
    [\ttt_{a+},\sss_{b+}]&=-\eta_{ab}\sss_{++} & [\ttt_{a+},\sss_{++}]&=0 & [\ttt_{a+},\sss_{cd}]&=-\eta_{ac}\sss_{+d}-\eta_{ad}\sss_{+c}\\
    [\ttt_{a+},\sss_{--}]&=2\sss_{a-} & [\ttt_{a+},\sss_{b-}]&=-\eta_{ab}\sss_{+-}+\sss_{ab} & [\ttt_{ab},\sss_{c+}]&=\eta_{ab}\sss_{a+}-\eta_{ac}\sss_{b+} \\
     [\ttt_{ab},\sss_{++}]&=0 & [\ttt_{ab},\sss_{--}]&=0 & [\ttt_{ab},\sss_{c-}]&=\eta_{bc}\sss_{a-}-\eta_{ac}\sss_{b-} \\
     [\ttt_{+-},\sss_{a+}]&=\sss_{+a} & [\ttt_{+-},\sss_{ab}]&=0 & [\ttt_{+-},\sss_{++}]&=2\sss_{++} \\
     [\ttt_{+-},\sss_{--}]&=-2\sss_{--} &[\ttt_{+-},\sss_{a-}]&=-\sss_{-a} & [\ttt_{a-},\sss_{b+}]&=-\eta_{ab}\sss_{-+}+\sss_{ab} \\
      [\ttt_{a-},\sss_{++}]&=2\sss_{a+} & [\ttt_{a-},\sss_{--}]&=0 &[\ttt_{a-},\sss_{cd}]&=-\eta_{ac}\sss_{-d}-\eta_{ad}\sss_{-c}  \\
     [\ttt_{a-},\sss_{b-}]&=-\eta_{ab}\sss_{--}&&&&
\end{align}
\esubeqs
\begin{align}
     [\ttt_{ab},\sss_{cd}]&=\eta_{bc}\sss_{ad}-\eta_{ac}\sss_{bd}+\eta_{bd}\sss_{ac}-\eta_{ad}\sss_{bc}\label{com2e3}
\end{align}
and for $[\SSS_{AB},\SSS_{AB}]$ 
\besubeqs
\begin{align}
    [\sss_{++},\sss_{a+}]&=0 & [\sss_{++},\sss_{ab}]&=0 & [\sss_{++},\sss_{++}]&=0 \\
    [\sss_{++},\sss_{--}]&=4\ttt_{+-} & [\sss_{++},\sss_{c-}]&=2\ttt_{+c} & [\sss_{a-},\sss_{c+}]&=\eta_{ac}\ttt_{-+}+\ttt_{ac}\\
    [\sss_{a-},\sss_{cd}]&=\eta_{ac}\ttt_{-d}+\eta_{ad}\ttt_{-c} & [\sss_{a-},\sss_{c-}]&=0& [\sss_{a-},\sss_{--}]&=0 
\end{align}\label{com3e3}
\esubeqs
\begin{align}
[\sss_{ab},\sss_{cd}]=\eta_{bc}\ttt_{ad}+\eta_{ac}\ttt_{bd}+\eta_{bd}\ttt_{ac}+\eta_{ad}\ttt_{bc}.\label{com4e3}
    \end{align}
The representation of the matrices in (\ref{com1e3}), (\ref{com2e3}),(\ref{com3e3}), (\ref{com4e3}) and (\ref{LPDK}) in PKLD basis with the $\eta_{ab}=diag(\pm1,1,1)$ is given in the Appendix E. We use the $\eta_{ab}=diag(1,1,1)$ in further calculations, and in the transformation map from PKLD basis to  WL basis. 

Similarly to the $sl(4)$ case we can define the near-horizon bcs by specifying $n=5$ in the general case (\ref{eq41}), and from (\ref{eq43}), we read out the gauge parameter $\epsilon$.
By defining $n=5$ we can also read out the charge from (\ref{eq46}) which is 
\begin{align}
    Q[\xi,\xi_3,\xi_4,\xi_5]=-\frac{k}{4\pi}\int d\phi&\left(\xi(\phi)\mak(\phi)+\alpha_3\xi_3(\phi)\mak_{0}^{(3)}(\phi)+\alpha_4\xi_4(\phi)\mak_{0}^{(4)}(\phi)\right.\nonumber\\& \left.+\alpha_5\xi_5(\phi)\mak_{0}^{(5)}(\phi) \right).
\end{align}
The algebra of charges $J_n,J_n^{(3)},J_n^{(4)}$ and $J_n^{(5)}$ defines ASA of the form (\ref{asaslg}) which corresponds to currents of the levels $\frac{1}{2}k$, $\frac{1}{2}\alpha_3k$,  $\frac{1}{2}\alpha_4 k$, and  $\frac{1}{2}\alpha_5k$.


To calculate the entropy we use $\Omega$ (\ref{eq41}) with $n=5$,  the definition for trace of $\langle L_0 L_0\rangle$ and $\langle W_0^{(s)}W_0^{(s)}\rangle$ matrices, and (\ref{entropygen}). 
This leads to 
    \begin{align}
    S=-\frac{k_N}{\pi}Im &\left[ \beta\int d\phi \left(\mak\mu_0\langle L_0L_0\rangle+\mak_{0}^{(3)}\mu_3\langle W_0^{(3)}W_0^{(3)}\rangle \right.\right.\nonumber\\ &\left.\left.+\mak_{0}^{(4)}\mu_4\langle W_0^{(4)}W_0^{(4)}\rangle+\mak_{0}^{(5)}\mu_5\langle W_0^{5}W_0^{(5)}\rangle\right)
    \right].
\end{align}
Using that $\langle W_0^{(s)}W_0^{(s)}\rangle=\frac{n(n^2-1)}{12}\alpha_s=n(n^2-1)\frac{4(s-1)!^4}{(2s-1)!(2s-2)!}\prod_{i=2}^{s-1}(n^2-i^2)$ we obtain 
    \begin{align}
    S=&-\frac{k_N}{\pi}Im\left[ \beta\int d\phi \left(10\mak\mu_0+56\mak_{0}^{(3)}\mu_3+\frac{288}{5}\mak_{0}^{(4)}\mu_4+\frac{1152}{35}\mak_{0}^{(5)}\mu_5\right)
    \right],\nonumber\\
    &=-2k_N\beta   \left(10\mak\mu_0+56\mak_{0}^{(3)}\mu_3+\frac{288}{5}\mak_{0}^{(4)}\mu_4+\frac{1152}{35}\mak_{0}^{(5)}\mu_5\right).
\end{align}
Similarly as in the $sl(4)$ and $so(3,2)$ case we calculate the chemical potentials to be 
\begin{align}
\mu_0=\frac{(p-k-2m-3n)\pi}{5\beta}, && \mu_3=\frac{3(n+p)-k}{14\beta}\pi, && \mu_4=\frac{(2k+4m+n+3p)\pi}{12\beta},\nonumber\\ 
\mu_5=\frac{5(2k+n+p)\pi}{24\beta}, && &&
\end{align}
for $p,k,n,m\in\mathbb{Z}$. 
 The entropy becomes 
\begin{align}
S&=-2k_N\pi\bigl[2(p-k-2m-3n)\mak+4\left(3(n+p)-k\right)\mak_{0}^{(3)}+\frac{24}{5}(2k+4m+n+3p)\mak_{0}^{(4)}\nonumber \bigr.\\&\bigl.+\frac{48}{7}(2k+n+p)\mak_{0}^{(5)} \bigr].
\end{align}
If we examine the map between the WL and PKLD matrices, we observe that each generator represented by a diagonal matrix in the WL basis contains the generator $\DDD$ and the generators $\SSLL_{ii}$, which are also represented by diagonal matrices. Since the generators $\SSLL_{ii}$ are not part of the Poincar\'e algebra and appear as linear combinations in each of the WL generators represented by diagonal matrices, this embedding cannot be used to compare the entropy in the $sl(4)$ and $so(3,2)$ cases. In these cases, only the field associated with the generator $\LLL_d$ in the bcs remains in the expressions for entropy, which are related to the BTZ entropy (as described in the paragraphs following equations (\ref{entropysl4}) and (\ref{entropyso32}), respectively).

We can, however, attempt a comparison with the $sl(4)$ scenario where the remaining field in the entropy related to BTZ entropy is associated with the generator $\LLL_0$ in the bcs (\ref{eq441}). For $sl(4)$ in this case, the entropy is $S = -\frac{20}{3}k_N\pi(1+2m)\mak$, whereas for the $sl(5)$ case, the entropy is $S = -40 k_N\pi m\mak$. This difference arises due to the distinct embeddings. In the $sl(5)$ scenario, the generator $\LLL_0$ is a linear combination of PKLD generators and is not proportional solely to the generator $\DDD$.

How the embedding of $so(3,2)$ affects the entropy becomes clear once we
recognize that different embeddings can describe different fields, such as
higher-spin fields or fields of another type. These different field contents
lead to different entropies. Since the choice of embedding determines which
fields remain in the theory, the entropy depends on the embedding as well. This was commented in the work \cite{Bunster:2014mua} in which principal and non-principal embeddings are studied. 

To consider metric formulation we need to determine the group element $b$.
We write the group element in terms of the generators $\PPP_i$ and a diagonal generator in the PKLD basis. If we consider $b=\rho \PPP_y+\PPP_t+\tilde{c}_1 \DDD$ we obtain degenerate dreibein. Reason is the representation of the dilatation generator.  
To obtain the example of the similar form as in $sl(4)$ and $so(3,2)$, we have to take the group element $b=a_2\rho \PPP_x+a_1 \PPP_y+a_3\rho \PPP_t+\SSS\LLL_{tt} $

The elements of dreibein obtained for that group element are $e_{\rho}^{\tilde{x}}=a_2$, $e_{\rho}^{\tilde{y}}=0,e_{\rho}^{\tilde{t}}=\frac{1}{2}a_3\sinh{2}$  and
\besubeqs\label{sl5dreibein}
\begin{align}
  e_i^{\tilde{x}}&=\frac{1}{70} a_2 \rho \left(35 X^i-210 X^i_3+252 X^i_4-120 X^i_5\right)\\
  e_{i}^{\tilde{y}}&=a_1\bigl(-\frac{X^i}{2}+3 X^i_3+\frac{12 X^i_5}{7}-\frac{18 X^i_4}
  {5}\bigr) \\ 
  e_i^{\tilde{t}}&=-\frac{\left(e^2-1\right)a_3\rho \left(35 \left(4 e^2-3\right) X^i-210 X^i_3+84 \left(1+2 e^2\right) X^i_4-120 X^i_5\right)}{140 e^2}. 
  \end{align}
  \esubeqs
  Here we denote $i=\{\phi,t\}$, for $i=\phi$ and $t$ respectively we have, $X^{\phi}=\mak, X_3^{\phi}=\mak_{0}^{(3)}, X^t=\mu, X_3^t=\mu_3$, etc..  An example which gives a similar dreibein  (\ref{dreisl5}) is analyzed in the Appendix E. 
The metric that dreibein (\ref{sl5dreibein}) leads to, is conformally flat, Einstein space, with Ricci scalar $R=-\frac{6}{a_1^2}$. The metric contains all the chemical potentials and fields, where fields can depend on the $\phi$ coordinate. For simplicity we set all chemical potentials and fields to zero besides $\mu_5$, $\mak_{0}^{(5)}$ and $\mak$ keeping them both constant. Such metric can be transformed to a form $ds^2=f_{rr}dr^2+f_{tt}dt^2+f_{t\phi}dtd\phi+r^2d\phi^2$ for the
\begin{footnotesize}
    \besubeqs \label{sl5ex}
\begin{align}
f_{rr}&=\frac{r^2}{\left(r^2-(\mak-\mak_{0}^{(5)})^2\right) \left(r^2+\left(\mak_{0}^{(5)}-\mak q_1\right){}^2\right)}\label{frrsl5}\\
f_{tt}&=\frac{ \left(-4 \mak q_2 \mak_{0}^{(5)} \left(\mak^2 q_1-\mak q_3 (2 \mak_{0}^{(5)}+\mu_5)+\mak_{0}^{(5)} (\mak_{0}^{(5)}+\mu_5)\right)+4 \mak^2 q_2^2 r^2+r^2 \left(-\mak q_3+\mak_{0}^{(5)}+\mu_5\right){}^2\right)}{4 \left(4 \mak^2 q_2^2+\left(\mak_{0}^{(5)}-\mak q_3\right){}^2\right)}\\
f_{t\phi}&=\frac{\mu_5 \left(2 \mak q_3-2 \mak_{0}^{(5)}\right) \left((\mak-\mak_{0}^{(5)})^2-r^2\right)}{\left(\mak_{0}^{(5)}-k q_1\right){}^2+(\mak-\mak_{0}^{(5)})^2}+\mu_5 (\mak_{0}^{(5)}-\mak)+r^2
\end{align}
\esubeqs
\end{footnotesize}
for $q_1=-3+4e^2,q_2=-1+e^2,q_3=-1+2e^2$. The parameters from the group element have been set to $a_1=1, a_2=\frac{1}{2e^2}$ and $a_3=\frac{1}{-1+e^2}$. The fields and the chemical potential have been suitably rescaled. The reason that we get the Euler's number is that in choosing the group element $b$ we allowed for the generators multiplied by constant terms, and not only those multiplied by function of $\rho$. If we kept function of $\rho$ we would obtain the terms with $e^{\rho}$ in the metric, which would lead to $\cosh(\rho)$ and $\sinh(\rho)$ functions. And we would have to consider the perturbative form of the Hausdorff formula. Keeping the dependence on $\rho$ only on selected places in the group element $b$ and allowing for the generators multiplied by constants, we obtain the terms multiplied by $e$ in the metric. 

We can see from (\ref{frrsl5}) that metric has one real and one imaginary eigenvalue, it belongs to $I_d$ class of one parameter subgroups of $SO(3,2)$ group \cite{Lovrekovic:2025dwd}. Setting the field $\mak_{0}^{(5)}$ to zero and rescaling the metric with -1 leads to  dS space where we have a metric with cosmological horizon.
Choosing the right parameter $b$ for the $sl(5)$ case, can also lead to Lobachevsky metric, which we describe in the Appendix E.

Here, we can also calculate the spin-3 field for the dreibein (\ref{sl5dreibein}) using\\ $\phi_{\mu\nu\rho}= e_{\mu}{}^ie_{\nu}{}^je_{\rho}{}^k \Tr[\TTT_i\TTT_j\TTT_k] $. 
We obtain \begin{align}
\varphi_{(\rho\phi t)}=-\frac{3}{70 e^4} \left[\left(e^2-1\right) \rho  \left(7 \mak_{0}^{(3)} \left(5 \mu +6 \mu _4\right)+4 \mak_{0}^{(5)} \left(5 \mu +6 \mu _4\right) \right.\right. \nonumber \\ \left.\left. -5 \mak \left(7 \mu _3-7 \mu _4+4 \mu _5\right)-\mak_{0}^{(4)} \left(35 \mu +42 \mu _3+24 \mu _5\right)\right)\right].
\end{align}
The brackets denote symmetry under permutation of indices. We use, for consistency, the Lorentz generators for generators over which a trace is taken. 
To be able to consistently treat the conformal scenario with higher spins and for the spin-2 we have to define the embedding of 
$so(3,2)$ in $sl(N)$ algebra so that we can continue with analogous treatment in higher-spin cases.

\section{Conclusion}

We have studied the holography of conformal HiSGRA theories by discussing the most general bcs and imposing near-horizon bcs.
For the holography of the most general bcs, we noted that the ASA corresponds to a loop algebra of the underlying gauge algebra. By selecting a specific set of generators from the most general bcs and retaining the chemical potentials, we defined near-horizon bcs, which led to an ASA corresponding to $u(1)$ currents.

Imposing the near-horizon bcs, we obtain solutions that can be categorized into classes studied under the classification of the one-parameter subgroups of 
$SO(3,2)$ \cite{Lovrekovic:2025dwd}. More precisely, once the near-horizon bcs are applied to the conformal spin-2 field, a suitable choice of the group element in the transformation from the first-order to the metric formulation can yield different solutions, depending on the choice of the projector $b$. We obtain solutions with:
(i) four real eigenvalues,
(ii) one real and one imaginary eigenvalue, which should belong to class $I_d$,
(iii) the Lobachevsky solution,
(iv) a solution that generalizes BTZ to the conformal scenario,
(v) a general solution that depends on three arbitrary functions of 
$\rho$, and
(vi) the simplest example, which reduces to BTZ.

We verify (iv) by solving the Cotton tensor equations of motion, as shown in Appendix B. A simplified form of metric (v) reproduces a 3D analog of the MKR solution.
For the near-horizon bcs of $sl(4)$, we present an example similar to solution (vi) of the conformal graviton. The modification arises due to the number of charges in the bcs. The distinction from the case when we have conformal spin-2 field appears when the group element $b$ includes a generator that commutes with the generator of translation, multiplied by an arbitrary function of $\rho$. This results in the appearance of an additional term in the metric.

As an example of the metric formulation for the near-horizon bcs of the conformal graviton and a spin-3 field defined from the $sl(5)$ algebra, we obtain a solution with one imaginary and one real eigenvalue, which fits the requirements for the solutions that should belong to the $I_d$ class in \cite{Lovrekovic:2025dwd}. For the same bcs, we also find a spin-3 field.

We calculate the entropies for the solutions coming from the near-horizon boundary conditions and look for the conditions under which they connect to known results.

\addtocontents{toc}{\protect\setcounter{tocdepth}{1}}

\section{Acknowledgements}
I would like to thank Daniel Grumiller and Evgeny Skvortsov for the  discussions, suggestions during the development of this work, and comments on the draft.
This work was supported by the Hertha Firnberg grant T 1269-N and Elise Richter grant V 1052-N of the Austrian Science Fund FWF.

\section{Appendix}
\subsection{Appendix A: Holography of Chern-Simons action}

Let us review the general approach to holography of the Chern-Simons action.

The action can be written in 2+1 decomposition as 
\begin{align}
    I=\frac{k}{4\pi} \int_{\mathbb{R}}dt\int_{\Sigma}d^2x\epsilon^{ij}g_{\mu\nu}(\dot{\omega}^{\mu}_i\omega_j^{\nu}+\omega_t^{\mu}F_{ij}^{\nu})+B(\partial\Sigma)
\end{align}
For $B$ the boundary term that depends on the bcs and needs to be included to ensure that the action is gauge invariant. The curvature $F_{ij}^{\mu}$ is $F_{ij}^{\mu}=\partial_{i}\omega_{j}^{\mu}-\partial_{j}\omega_i^{\mu}+f^{\mu}_{\nu\rho}\omega^{\nu}\omega^{\rho}$
and the $f^{\mu}_{\nu\rho}$ are the structure constants of the Lie algebra. 
The indices $\mu,\nu$ are raised and lowered with the metric $g_{\mu\nu}$ which is invertible.
$\omega_t^{\mu}$ is treated as a Lagrange mutliplier and the fields $\omega_i^{\mu}$ are dynamical fields which satisfy Poisson brackets. The variation of the action with respect to the $\omega_{t}^{\mu}$ gives the constraint $G_{\mu}\equiv\frac{k}{4\pi}g_{\mu\nu}\epsilon^{ij}F_{ij}^{\nu}$ which satisfies Poisson bracket algebra
\begin{align}
    \{ G_{\mu}(x),G_{\nu}(y)\}=f^{\rho}_{\mu\nu}G_{\rho}(x)\delta(x,y)
\end{align}
so $G_{\mu}$ are generators of gauge transformations which act on the phase space. 
The smeared generators $G(\xi^{\mu})=\int_{\Sigma}\xi^{\mu}G_{\mu}+Q(\xi)$
give central extension of the algebra of the global charges.
We assume that the parameter $\xi$ does not depend on the $\omega_{i}$ in the interior, such that the variation of the smeared generator leads to 
\begin{align}
    \delta G=\frac{k}{2\pi}\int_{\Sigma}\epsilon^{ij}\xi_{\mu}D_i\delta\omega_j^{\mu}+\delta Q(\xi)
\end{align}
for $D_i v^{\mu}\equiv\partial_i v^{\mu}+f^{\mu}_{\nu\rho}\omega_i^{\nu}v^{\rho}$ covariant derivative. Here, $v^{\mu}$ is an arbitrary vector field. Integration by parts of the first term leads to the 
\begin{align}
    \delta G=-\frac{k}{2\pi}\int_{\Sigma}\epsilon^{ij}D_i\xi_{\mu}\delta\omega_j^{\nu}+\frac{k}{2\pi}\int_{\partial\Sigma} \xi_{\mu}\delta\omega_k^{\mu}dx^k +\delta Q.
\end{align}
The functional derivative of $G$ then becomes well defined and the surface terms cancel out if we demand that the variation of the charge is given by 
\begin{align}
    \delta Q=-\frac{k}{2\pi} \int_{\partial \Sigma}\xi_{\mu}\delta \omega_i^{\mu}dx^i.\label{varch}
\end{align}
The Poisson bracket of the two smeared generators, with the parameters $\xi$ and $\lambda$
\begin{align}
    \{G(\xi),G(\lambda) \}=-\frac{k}{2\pi}\int_{\Sigma}D\xi_a\wedge D \lambda^a
\end{align}
can be integrated by parts to obtain 
\begin{align}
    \{G(\xi),G(\lambda) \}=\int_{\Sigma}[\xi,\lambda]^aG_a+\frac{k}{2\pi}\int_{\partial\Sigma}\xi_a D\lambda^a\label{pg}
\end{align}
For $[\xi,\lambda]^a=f^a_{bc}\xi^b\lambda^c$ commutator of the algebra. 

To obtain the smeared generator  one needs to add the surface term to the first term on the right to make it differentiable. That means the second term on the right hand side, the surface term, needs be the same as the charge which regularizes the first term up to a central extension. Therefore, one can write \begin{align}
    \frac{k}{2\pi}\int \xi_aD\lambda^a=Q(\sigma)+K(\xi,\lambda)\label{che}
\end{align}
for the  Q defined in (\ref{varch}), and $\sigma$  a function of $\lambda$ and $\xi$. Its precise form will define the algebra of the global charges, which depends on the bcs. The second term on the right hand side of (\ref{che}) is the central extension and it does not depend on the variables varied on the boundary.

Combining the (\ref{che}), (\ref{pg})  replacing the constraints with their charges and promoting the Poisson bracket into Dirac bracket  one obtains 
\begin{align}
    \{Q(\lambda),Q(\xi)\}^*=Q(\sigma)+K(\xi,\lambda).
\end{align}
We will consider the simple bcs for which one can integrate the above expression (\ref{varch}), and parameter $\xi_{\mu}$ depends linearly on the fields
\begin{align}
    \xi^{i}=-\xi^{i}_{\mu}\omega^{\mu}.
\end{align}

\subsubsection{Gauge fixing}

 The bcs that we consider here, require choosing the radial component as \begin{align} \omega_{\rho}=b(\rho)^{-1}\partial_{\rho}b(\rho)  \end{align} where group element $b$ depends only on radial coordinate. This has been referred to as partial fixing to radial gauge \cite{Grumiller:2016kcp} and it has been standardly used in the analysis of Chern-Simons action because it simplifies the later analysis, and it is also a choice that is always possible \cite{Campoleoni:2010zq}.
 In both cases of bcs that we impose, we use \begin{align}
 \omega=b(\rho)^{-1}(d+\Omega(t,\phi))b(\rho) \label{bcsx}
 \end{align}
form of the gauge field. 
\subsection{Appendix B}

Here, we compare our solution for  conformal spin-2 field in example 3, to most general solution of the Cotton tensor, of the form $ds^2=\frac{dr^2}{f(r)}+(r^2n(r,t)^2-f(r))dt^2-2r^2n(r,t)dtd\phi+r^2d\phi^2$. 
The solution reads 
\begin{footnotesize}
    \besubeqs
\begin{align}
f(r)&=c_3 r^2+\frac{c_2{}^2}{4 r^2}+c_1 \left(\frac{c_2}{r}-c_4 r\right)+c_1{}^2-\frac{c_2 c_4}{2}\\
f(t)&=r^2n(r,t)^2-f(r)\\&=\left(c_3{}^2+(-1+2 c_5) c_3+c_5{}^2\right) r^2+c_1 (-2 c_3+c_4-2 c_5) r+\frac{1}{2} c_2 (-2 c_3+c_4-2 c_5)\\
n(r,t)&=\frac{1}{2r^2}(-2 (c_3+c_5) r^2+2 c_1 r+c_2) 
\end{align}
\esubeqs
\end{footnotesize}
and it is equivalent to the solution of the example 2 for the 
\begin{footnotesize}
\besubeqs
\begin{align}
c_1&\to -2\mak\mak_d q_1\\
c_2&\to 2 \mak_d \mak \\
c_3&\to \mak^2 \mak_d^2 q_1^4-\mak^2 q_1^2-\mak_d^2 q_1^2+1\\
c_4&\to -\frac{\mak^2+\mak_d^2-2\mak^2\mak_d^2 q_1^2}{\mak\mak_d} \\
c_5&\to -\mak^2 \mak_d^2 q_1^4+q_1^2 \left(\mak^2-\mak \mak_d+\mak_d^2\right)+\frac{(\mak-\mak_d) (\mu+\mu_d)}{\mak_d \mu-\mak \mu_d}+q_2.
\end{align}
\esubeqs
\end{footnotesize}
Here, the function $f(r)$ agrees with the one from the example 2 for $\frac{\sqrt{2 \mak^2+\mak_d^2+\mak_d}}{2 \mak \mak_d} $. The function $f(r)$ has null points for the 
\begin{footnotesize}
    \besubeqs
\begin{align}
r_{1\pm}\to \frac{c_1 \left(c_4-\sqrt{c_4{}^2-4 c_3}\right)\pm\sqrt{2 \left(c_4{}^2-\sqrt{c_4{}^2-4 c_3} c_4-2 c_3\right) c_1{}^2+4 c_2 c_3 \left(c_4-\sqrt{c_4{}^2-4 c_3}\right)}}{4 c_3}\\
r_{2\pm}\to \frac{c_1 \left(c_4+\sqrt{c_4{}^2-4 c_3}\right)\pm\sqrt{2 \left(c_4 \left(c_4+\sqrt{c_4{}^2-4 c_3}\right)-2 c_3\right) c_1{}^2+4 c_2 c_3 \left(c_4+\sqrt{c_4{}^2-4 c_3}\right)}}{4 c_3}
\end{align}
\esubeqs
\end{footnotesize}
which shows that one obtained the solution which can have complex eigenvalues, depending on the coefficients $c_1,c_2,...,c_5$. That makes it belong to type $I_a$ from the classification according to one parameter subgroups of $SO(3,2)$ group \cite{Lovrekovic:2025dwd}.

\subsection{Appendix C}

{\bf Metric formulation -- example 3.} In this example we analyze how changing the dependence on $\rho$ in front of a generator in the projector $b$ influences the metric. We set $a(\rho)$ and $b(\rho)$ in (\ref{gengen}) to one, rescale the metric with $(-1+e)$, and then consider it while keeping only $c(\rho)$. %

The Ricci scalar of this metric is $-2$ which shows that the metric can be brought to more familiar Lobachevsky form. By transformation of the coordinates, rescaling of the fields one obtains the metric of the Lobachevsky form \cite{Bertin:2012qw}
\begin{align}
ds^2=\frac{dr^2}{r^2-Br+A^2+C}-dt^2+2A dtd\phi+(r^2-Br)d\phi^2. \label{lobpaper}
\end{align}
 If we  use the function $c(\rho)$ so that $c(\rho)=q_2+q_1\rho$ we can transform $t$ and $\phi$ coordinates into $\tilde{t}$ and $\tilde{\phi}$ respectively (however in the metric we drop the $\sim$ to avoid clutter), to set $g_{\rho\phi}=g_{\rho t}=0$. The metric (\ref{gengen}) then takes the form 
 \besubeqs
\begin{align}
ds^2&=\frac{\mak^2 q_1^2 }{\mak q_1\rho \left(2 \sqrt{\mak_d^2-\mak^2}+\mak q_1 \rho\right)+\mak_d^2}d\rho ^2-dt^2 +\frac{2 \mak_d }{\mak q_1}dtd\phi\nonumber \\&+\left(\frac{2 \rho  \sqrt{\mak_d^2-k^2}}{\mak q_1}+\rho ^2\right)d\phi^2.\label{lobmet}
  \end{align}
\esubeqs
In this notation we have $B=-\frac{2   \sqrt{\mak_d^2-\mak^2}}{\mak q_1}$ and $A=\frac{ \mak_d }{\mak q_1}$, for $C=0$.
Comparing the solution to (\ref{lobpaper}), for $q_1=1$ and: $\mak_d=0$ we recover global Lobachevsky; for $\mak_d=\mak=0$ we obtain Poincar\'e Lobachevsky; and for $\mak=\mak_d$ rotating Lobachevsky metric. 

If we choose the $a(\rho)$ so that $\frac{\left(e^{a(\rho)}-1\right)^2}{a(\rho)^2}$ in front of the generator of dilatations, $\DDD$, is a constant, the Ricci scalar gets rescaled by that factor. 
\\
\indent Interesting to notice is that starting from  $Y=b(\rho)\PPP_y+\DDD+c(\rho)\PPP_x$ and $\mu=0$ the dreibein becomes combination of above (\ref{lobmet}) example, and the simplest example 1.
One obtains the Ricci scalar as a function of the $b(\rho)$ and $c(\rho)$. The choice $b(\rho)=1$ and $c(\rho)$ free as in the previous example, leads to the metric that has Ricci scalar $-\frac{2}{c_1^2}$ while the choice $c(\rho)=1$ and $b(\rho)$ free leads to $R=-\frac{6}{c_1^2}$ Ricci scalar, as in the example 1. \\ 

\subsection{Appendix D}

The $sl(n)$ generators are in our convention represented by NxN matrices 
\begin{align}
   (\LLL_1)_{jk}=-\sqrt{j(N-j)}\delta_{j+1,k}\\
   (\LLL_{-1})_{jk}=\sqrt{k(N-k)}\delta_{j,k+1}\\
   (\LLL_0)_{jk}=\frac{1}{2}(N+1-2j)\delta_{j,k}\end{align}
where $j,k=1,..,N$ ($j+1,k+1=2,..N+1$) while the remaining $\WWW$ generators are defined by 
\begin{align}
    \WWW_m^{(s)}=2(-1)^{s-m-1}\frac{(s+m-1)!}{(2s-2)!}[\LLL_{-1},[\LLL_{-1}...[\LLL_{-1},(\LLL_1)^{s-1}]..]] \label{eq34}
\end{align}
with $s-m-1$ as a number of commutations. This basis we call WL basis.

In following we define $\VVV_m^{4}\equiv \WWW^4_m$ and $\VVV_n^{3}\equiv \WWW^3_n$. 
 With the help of basis of generators for the principal embedding of spin-4 gravity \cite{Grumiller:2016kcp}, $sl(2)$ generators
\begin{align}
    \LLL_0&=\frac{1}{2}\left(\begin{array}{cccc}
         3&0&0&0  \\
         0&1&0&0 \\
         0&0&-1&0\\
         0&0&0&-3
    \end{array}\right), &&\LLL_{-1}=\left(\begin{array}{cccc}
         0&0&0&0  \\
         \sqrt{3}&0&0&0 \\
         0&2&0&0\\
         0&0&\sqrt{3}&0
    \end{array}\right),&&\LLL_1=\left(
\begin{array}{cccc}
 0 & -\sqrt{3} & 0 & 0 \\
 0 & 0 & -2 & 0 \\
 0 & 0 & 0 & -\sqrt{3} \\
 0 & 0 & 0 & 0 \\
\end{array}
\right), \label{wl41}
\end{align}
quintet 
\begin{align}
    \VVV^{(3)}_0&=\left(
\begin{array}{cccc}
 2 & 0 & 0 & 0 \\
 0 & -2 & 0 & 0 \\
 0 & 0 & -2 & 0 \\
 0 & 0 & 0 & 2 \\
\end{array}
\right), &&\VVV^{(3)}_1=\left(
\begin{array}{cccc}
 0 & 2 \sqrt{3} & 0 & 0 \\
 0 & 0 & 0 & 0 \\
 0 & 0 & 0 & -2 \sqrt{3} \\
 0 & 0 & 0 & 0 \\
\end{array}
\right),&&\VVV^{(3)}_{-1}=\left(
\begin{array}{cccc}
 0 & 0 & 0 & 0 \\
 -2 \sqrt{3} & 0 & 0 & 0 \\
 0 & 0 & 0 & 0 \\
 0 & 0 & 2 \sqrt{3} & 0 \\
\end{array}
\right),\nonumber \\
     \VVV^{(3)}_2&=\left(
\begin{array}{cccc}
 0 & 0 & 4 \sqrt{3} & 0 \\
 0 & 0 & 0 & 4 \sqrt{3} \\
 0 & 0 & 0 & 0 \\
 0 & 0 & 0 & 0 \\
\end{array}
\right), &&\VVV^{(3)}_{-2}=\left(
\begin{array}{cccc}
 0 & 0 & 0 & 0 \\
 0 & 0 & 0 & 0 \\
 4 \sqrt{3} & 0 & 0 & 0 \\
 0 & 4 \sqrt{3} & 0 & 0 \\
\end{array}
\right)&&\label{wl42}\end{align}
and septet
\begin{align}
         \VVV^{(4)}_2&=\left(
\begin{array}{cccc}
 0 & 0 & -2 \sqrt{3} & 0 \\
 0 & 0 & 0 & 2 \sqrt{3} \\
 0 & 0 & 0 & 0 \\
 0 & 0 & 0 & 0 \\
\end{array}
\right), &&\VVV^{(4)}_{-2}=\left(
\begin{array}{cccc}
 0 & 0 & 0 & 0 \\
 0 & 0 & 0 & 0 \\
 -2 \sqrt{3} & 0 & 0 & 0 \\
 0 & 2 \sqrt{3} & 0 & 0 \\
\end{array}
\right),&&\nonumber \\
    \VVV^{(4)}_0&=\left(
\begin{array}{cccc}
 -\frac{3}{5} & 0 & 0 & 0 \\
 0 & \frac{9}{5} & 0 & 0 \\
 0 & 0 & -\frac{9}{5} & 0 \\
 0 & 0 & 0 & \frac{3}{5} \\
\end{array}
\right),  &&\VVV^{(4)}_1=\frac{-4}{5}\left(
\begin{array}{cccc}
 0 & \sqrt{3} & 0 & 0 \\
 0 & 0 & -3 & 0 \\
 0 & 0 & 0 &  \sqrt{3} \\
 0 & 0 & 0 & 0 \\
\end{array}
\right),&&\VVV^{(4)}_{-1}=\frac{4}{5}\left(
\begin{array}{cccc}
 0 & 0 & 0 & 0 \\
  \sqrt{3} & 0 & 0 & 0 \\
 0 & -3 & 0 & 0 \\
 0 & 0 &  \sqrt{3} & 0 \\
\end{array}
\right), \nonumber\\
\VVV^{(4)}_{-3}&=\left(\begin{array}{cccc}
         0&0&0&0  \\
         0&0&0&0 \\
         0&0&0&0\\
         12&0&0&0
    \end{array}\right), &&\VVV^{(4)}_{3}=\left(\begin{array}{cccc}
         0&0&0&-12  \\
         0&0&0&0 \\
         0&0&0&0\\
         0&0&0&0
    \end{array}\right),&&\label{wl43}
    \end{align}
and  generators  in our PKLD basis
\begin{align}
\PPP_t&=\left(
\begin{array}{cccc}
 0 & 0 & -\frac{1}{2} & 0 \\
 0 & 0 & 0 & \frac{1}{2} \\
 0 & 0 & 0 & 0 \\
 0 & 0 & 0 & 0 \\
\end{array}
\right),\PPP_y=\left(
\begin{array}{cccc}
 0 & 0 & \frac{1}{2} & 0 \\
 0 & 0 & 0 & \frac{1}{2} \\
 0 & 0 & 0 & 0 \\
 0 & 0 & 0 & 0 \\
\end{array}
\right),\PPP_x=\left(
\begin{array}{cccc}
 0 & 0 & 0 & 1 \\
 0 & 0 & -\frac{1}{4} & 0 \\
 0 & 0 & 0 & 0 \\
 0 & 0 & 0 & 0 \\
\end{array}
\right),\DDD=\left(
\begin{array}{cccc}
 -\frac{1}{2} & 0 & 0 & 0 \\
 0 & -\frac{1}{2} & 0 & 0 \\
 0 & 0 & \frac{1}{2} & 0 \\
 0 & 0 & 0 & \frac{1}{2} \\
\end{array}
\right)   \nonumber \\ \KKK_t&=\left(
\begin{array}{cccc}
 0 & 0 & 0 & 0 \\
 0 & 0 & 0 & 0 \\
 -1 & 0 & 0 & 0 \\
 0 & 1 & 0 & 0 \\
\end{array}
\right),\KKK_y=\left(
\begin{array}{cccc}
 0 & 0 & 0 & 0 \\
 0 & 0 & 0 & 0 \\
 -1 & 0 & 0 & 0 \\
 0 & -1 & 0 & 0 \\
\end{array}
\right),\KKK_x=\left(
\begin{array}{cccc}
 0 & 0 & 0 & 0 \\
 0 & 0 & 0 & 0 \\
 0 & 2 & 0 & 0 \\
 -\frac{1}{2} & 0 & 0 & 0 \\
\end{array}
\right),\LLL_{ty}=\left(
\begin{array}{cccc}
 -\frac{1}{2} & 0 & 0 & 0 \\
 0 & \frac{1}{2} & 0 & 0 \\
 0 & 0 & \frac{1}{2} & 0 \\
 0 & 0 & 0 & -\frac{1}{2} \\
\end{array}
\right)\nonumber \\ \LLL_{tx}&=\left(
\begin{array}{cccc}
 0 & 1 & 0 & 0 \\
 \frac{1}{4} & 0 & 0 & 0 \\
 0 & 0 & 0 & 1 \\
 0 & 0 & \frac{1}{4} & 0 \\
\end{array}
\right),\LLL_{yx}=\left(
\begin{array}{cccc}
 0 & -1 & 0 & 0 \\
 \frac{1}{4} & 0 & 0 & 0 \\
 0 & 0 & 0 & 1 \\
 0 & 0 & -\frac{1}{4} & 0 \\
\end{array}
\right),\ttt_+=\left(
\begin{array}{cccc}
 0 & 0 & 0 & -1 \\
 0 & 0 & -\frac{1}{4} & 0 \\
 0 & 0 & 0 & 0 \\
 0 & 0 & 0 & 0 \\
\end{array}
\right),\ttt_-=\left(
\begin{array}{cccc}
 0 & 0 & 0 & 0 \\
 0 & 0 & 0 & 0 \\
 0 & -2 & 0 & 0 \\
 -\frac{1}{2} & 0 & 0 & 0 \\
\end{array}
\right) \nonumber \\
\ttt_t&=\left(
\begin{array}{cccc}
 0 & 1 & 0 & 0 \\
 -\frac{1}{4} & 0 & 0 & 0 \\
 0 & 0 & 0 & 1 \\
 0 & 0 & -\frac{1}{4} & 0 \\
\end{array}
\right),\ttt_y=\left(
\begin{array}{cccc}
 0 & -1 & 0 & 0 \\
 -\frac{1}{4} & 0 & 0 & 0 \\
 0 & 0 & 0 & 1 \\
 0 & 0 & \frac{1}{4} & 0 \\
\end{array}
\right),\ttt_x=\left(
\begin{array}{cccc}
 -\frac{1}{2} & 0 & 0 & 0 \\
 0 & \frac{1}{2} & 0 & 0 \\
 0 & 0 & -\frac{1}{2} & 0 \\
 0 & 0 & 0 & \frac{1}{2} \\
\end{array}
\right) \label{matex1}
\end{align}
which closes algebra of the vector field and conformal spin-2, for the $\eta_{ab}=diag(-1,1,1)$ for (t,y,x), we can determine
the map 
\besubeqs
\begin{footnotesize}
\begin{align}
\LLL_{-1}&\to 2 \left(\frac{1}{4} \KKK_x+\sqrt{3} \left(\LLL_{tx}-\ttt_t\right)-\frac{1}{4} \ttt_-\right), && \LLL_0\to  \left(2 \DDD+\ttt_x\right), \\ 
\LLL_1&\to \frac{1}{2} \left( -\sqrt{3} \LLL_{tx}+8\PPP_x+8\ttt_+- \sqrt{3} \ttt_t\right), && \VVV^{(3)}{}_{-2}\to -4 \sqrt{3}\KKK_y,\\
\VVV^{(3)}{}_0&\to -4\LLL_{ty}, && \VVV^{(3)}{}_1\to -\sqrt{3}(\LLL_{yx}+\ttt_y),\\
\VVV^{(3)}{}_2&\to 8 \sqrt{3}\PPP_y,&& \VVV^{(4)}{}_{-3}\to -12 \left(\KKK_x+\ttt_-\right), \\
\VVV^{(4)}{}_{-2}&\to 2 \sqrt{3}\KKK_t, && \VVV^{(4)}{}_{-1}\to \frac{1}{5} \left(-3 \KKK_x+8\sqrt{3} \left(\LLL_{yx}-\ttt_x\right)+3 \ttt_-\right),\\
\VVV^{(4)}{}_0&\to \frac{6}{5} \left(2 \ttt_x-\DDD\right), && \VVV^{(4)}{}_1\to \frac{2}{5} \left(- \sqrt{3} \LLL_{tx}-12 \PPP_x-12 \ttt_+- \sqrt{3} \ttt_t\right),\\
\VVV^{(4)}{}_2 &\to 4 \sqrt{3}\PPP_t, && \VVV^{(4)}{}_3\to 6 \left(\ttt_+-\PPP_x\right) \\ \VVV^{(3)}{}_{-1}&\to 4 \sqrt{3}(\ttt_y-\LLL_{yx}). && 
\end{align}
\end{footnotesize}\label{mapsl4}
\esubeqs
The matrices (\ref{matex1}) are related to standard construction using $\gamma$ matrices  $\gamma_1,\gamma_2,\gamma_3,\gamma_4,\gamma_5$ in Majorana basis via following. We obtain 9 linearly independent matrices for 
\begin{align}
    v_1&=[\gamma_1,\gamma_4]/2, && v_2=[\gamma_2,\gamma_3]/2, && v_3=[\gamma_5,\gamma_1]/2, && v_4=[\gamma_1,\gamma_3]/2, &&
    v_5=[\gamma_2,\gamma_4]/2, \nonumber\\ v_6&=[\gamma_2,\gamma_5]/2,  &&v_7=[\gamma_1,\gamma_2]/2, && v_8=[\gamma_3,\gamma_4]/2, &&
    v_9=[\gamma_3,\gamma_5]/2, 
\end{align}
multiplying which we construct additional 6 ones
\begin{align}
    v_{10}&=v_7.v_9, && v_{11}=v_7.v_8, && v_{12}=v_5.v_6, &&v_{13}=v_2.v_{12}, && v_{14}=v_4.v_{12}, && v_{15}=v_7.v_{12}.
\end{align}
Together they define the (\ref{matex1})
\besubeqs
\begin{align}
    \PPP_t&=\frac{v_1}{4}-\frac{v_8}{4}, && \PPP_y=\frac{v_3}{4}+\frac{v_9}{4}&& \PPP_x=\frac{1}{16} \left(-3 v_2-3 v_7-5 \left(v_{13}+v_{15}\right)\right)\\
    \DDD&=-\frac{v_4}{2}, && \KKK_t=\frac{1}{2} \left(v_1+v_8\right) && \KKK_x=\frac{1}{8} \left(3 v_2-3 v_7-5 v_{13}+5 v_{15}\right) \\
    \KKK_y&=\frac{1}{2} \left(v_3-v_9\right) && \LLL_{ty}=-\frac{v_{12}}{2}&& \ttt_+=\frac{1}{16} \left(5 v_2+5 v_7+3 \left(v_{13}+v_{15}\right)\right) \\
    \LLL_{tx}&=\frac{1}{8} \left(3 v_5-5 v_{10}\right) && \LLL_{yx}= \frac{1}{8} \left(5 v_{11}-3 v_6\right) && \ttt_-=\frac{1}{8} \left(-5 v_2+5 v_7+3 v_{13}-3 v_{15}\right) \\
    \ttt_t&=\frac{1}{8} \left(5 v_5-3 v_{10}\right) && \ttt_y=\frac{1}{8} \left(3 v_{11}-5 v_6\right) && \ttt_x=-\frac{v_{14}}{2}
    \end{align}
    \esubeqs

\subsection{Appendix E}

Using the identification $\sss_{ab}\equiv \SSLL_{ab},\sss_{a+}\equiv \SP_a,\sss_{a-}\equiv \SK_a,\SSDD=\sss_{+-}$ the matrices that satisfy the commutation relations (\ref{com1e3}), (\ref{com2e3}),(\ref{com3e3}), (\ref{com4e3}) and (\ref{LPDK}) for the $\eta_{ab}=diag(\pm1,1,1)$ can be written as:
\begin{align} \PPP_x&=\left(
\begin{array}{ccccc}
 0 & 0 & \mp1 & 0 & 0 \\
 0 & 0 & 0 & 0 & 0 \\
 0 & 1 & 0 & 0 & 0 \\
 0 & 0 & 0 & 0 & 0 \\
 0 & 0 & 0 & 0 & 0 \\
\end{array}
\right),\PPP_y=\left(
\begin{array}{ccccc}
 0 & 0 & 0 & -1 & 0 \\
 0 & 0 & 0 & 0 & 0 \\
 0 & 0 & 0 & 0 & 0 \\
 0 & 1 & 0 & 0 & 0 \\
 0 & 0 & 0 & 0 & 0 \\
\end{array}
\right),\PPP_t=\left(
\begin{array}{ccccc}
 0 & 0 & 0 & 0 & -1 \\
 0 & 0 & 0 & 0 & 0 \\
 0 & 0 & 0 & 0 & 0 \\
 0 & 0 & 0 & 0 & 0 \\
 0 & 1 & 0 & 0 & 0 \\
\end{array}
\right),\nonumber \\ \KKK_x&=\left(
\begin{array}{ccccc}
 0 & 0 & 0 & 0 & 0 \\
 0 & 0 & \mp1 & 0 & 0 \\
 1 & 0 & 0 & 0 & 0 \\
 0 & 0 & 0 & 0 & 0 \\
 0 & 0 & 0 & 0 & 0 \\
\end{array}
\right), \KKK_y=\left(
\begin{array}{ccccc}
 0 & 0 & 0 & 0 & 0 \\
 0 & 0 & 0 & -1 & 0 \\
 0 & 0 & 0 & 0 & 0 \\
 1 & 0 & 0 & 0 & 0 \\
 0 & 0 & 0 & 0 & 0 \\
\end{array}
\right), \KKK_t=\left(
\begin{array}{ccccc}
 0 & 0 & 0 & 0 & 0 \\
 0 & 0 & 0 & 0 & -1 \\
 0 & 0 & 0 & 0 & 0 \\
 0 & 0 & 0 & 0 & 0 \\
 1 & 0 & 0 & 0 & 0 \\
\end{array}
\right),\DDD=-\left(
\begin{array}{ccccc}
 1 & 0 & 0 & 0 & 0 \\
 0 & -1 & 0 & 0 & 0 \\
 0 & 0 & 0 & 0 & 0 \\
 0 & 0 & 0 & 0 & 0 \\
 0 & 0 & 0 & 0 & 0 \\
\end{array}
\right),\nonumber\\ \LLL_{xy}&=\left(
\begin{array}{ccccc}
 0 & 0 & 0 & 0 & 0 \\
 0 & 0 & 0 & 0 & 0 \\
 0 & 0 & 0 & 1 & 0 \\
 0 & 0 & \mp1 & 0 & 0 \\
 0 & 0 & 0 & 0 & 0 \\
\end{array}
\right), \LLL_{xt}=\left(
\begin{array}{ccccc}
 0 & 0 & 0 & 0 & 0 \\
 0 & 0 & 0 & 0 & 0 \\
 0 & 0 & 0 & 0 & 1 \\
 0 & 0 & 0 & 0 & 0 \\
 0 & 0 & \mp1 & 0 & 0 \\
\end{array}
\right), \LLL_{yt}=\left(
\begin{array}{ccccc}
 0 & 0 & 0 & 0 & 0 \\
 0 & 0 & 0 & 0 & 0 \\
 0 & 0 & 0 & 0 & 0 \\
 0 & 0 & 0 & 0 & 1 \\
 0 & 0 & 0 & -1 & 0 \\
\end{array}
\right)\label{pkl5}
\end{align}

\begin{align}
   \SP_x=& \left(
\begin{array}{ccccc}
 0 & 0 & \pm1 & 0 & 0 \\
 0 & 0 & 0 & 0 & 0 \\
 0 & 1 & 0 & 0 & 0 \\
 0 & 0 & 0 & 0 & 0 \\
 0 & 0 & 0 & 0 & 0 \\
\end{array}
\right),\SP_y=\left(
\begin{array}{ccccc}
 0 & 0 & 0 & 1 & 0 \\
 0 & 0 & 0 & 0 & 0 \\
 0 & 0 & 0 & 0 & 0 \\
 0 & 1 & 0 & 0 & 0 \\
 0 & 0 & 0 & 0 & 0 \\
\end{array}
\right),\SP_t=\left(
\begin{array}{ccccc}
 0 & 0 & 0 & 0 & 1 \\
 0 & 0 & 0 & 0 & 0 \\
 0 & 0 & 0 & 0 & 0 \\
 0 & 0 & 0 & 0 & 0 \\
 0 & 1 & 0 & 0 & 0 \\
\end{array}
\right),\SK_x=\left(
\begin{array}{ccccc}
 0 & 0 & 0 & 0 & 0 \\
 0 & 0 & \pm1 & 0 & 0 \\
 1 & 0 & 0 & 0 & 0 \\
 0 & 0 & 0 & 0 & 0 \\
 0 & 0 & 0 & 0 & 0 \\
\end{array}
\right),\nonumber\\ \SK_y&=\left(
\begin{array}{ccccc}
 0 & 0 & 0 & 0 & 0 \\
 0 & 0 & 0 & 1 & 0 \\
 0 & 0 & 0 & 0 & 0 \\
 1 & 0 & 0 & 0 & 0 \\
 0 & 0 & 0 & 0 & 0 \\
\end{array}
\right),\SK_t=\left(
\begin{array}{ccccc}
 0 & 0 & 0 & 0 & 0 \\
 0 & 0 & 0 & 0 & 1 \\
 0 & 0 & 0 & 0 & 0 \\
 0 & 0 & 0 & 0 & 0 \\
 1 & 0 & 0 & 0 & 0 \\
\end{array}
\right),\SSDD=\left(
\begin{array}{ccccc}
 \frac{3}{5} & 0 & 0 & 0 & 0 \\
 0 & \frac{3}{5} & 0 & 0 & 0 \\
 0 & 0 & -\frac{2}{5} & 0 & 0 \\
 0 & 0 & 0 & -\frac{2}{5} & 0 \\
 0 & 0 & 0 & 0 & -\frac{2}{5} \\
\end{array}
\right),\SSLL_{xy}=\left(
\begin{array}{ccccc}
 0 & 0 & 0 & 0 & 0 \\
 0 & 0 & 0 & 0 & 0 \\
 0 & 0 & 0 & 1 & 0 \\
 0 & 0 & \pm1 & 0 & 0 \\
 0 & 0 & 0 & 0 & 0 \\
\end{array}
\right),\nonumber\end{align}
\begin{align}\SSLL_{xt}&=\left(
\begin{array}{ccccc}
 0 & 0 & 0 & 0 & 0 \\
 0 & 0 & 0 & 0 & 0 \\
 0 & 0 & 0 & 0 & 1 \\
 0 & 0 & 0 & 0 & 0 \\
 0 & 0 & \pm1 & 0 & 0 \\
\end{array}
\right),\SSLL_{yt}=\left(
\begin{array}{ccccc}
 0 & 0 & 0 & 0 & 0 \\
 0 & 0 & 0 & 0 & 0 \\
 0 & 0 & 0 & 0 & 0 \\
 0 & 0 & 0 & 0 & 1 \\
 0 & 0 & 0 & 1 & 0 \\
\end{array}
\right),\sss_{++}=\left(
\begin{array}{ccccc}
 0 & 2 & 0 & 0 & 0 \\
 0 & 0 & 0 & 0 & 0 \\
 0 & 0 & 0 & 0 & 0 \\
 0 & 0 & 0 & 0 & 0 \\
 0 & 0 & 0 & 0 & 0 \\
\end{array}
\right),\sss_{--}=\left(
\begin{array}{ccccc}
 0 & 0 & 0 & 0 & 0 \\
 2 & 0 & 0 & 0 & 0 \\
 0 & 0 & 0 & 0 & 0 \\
 0 & 0 & 0 & 0 & 0 \\
 0 & 0 & 0 & 0 & 0 \\
\end{array}
\right),\nonumber\\\SSLL_{xx}&=\left(
\begin{array}{ccccc}
 \mp\frac{2}{5} & 0 & 0 & 0 & 0 \\
 0 & \mp\frac{2}{5} & 0 & 0 & 0 \\
 0 & 0 & \pm\frac{8}{5} & 0 & 0 \\
 0 & 0 & 0 & \mp\frac{2}{5} & 0 \\
 0 & 0 & 0 & 0 & \mp\frac{2}{5} \\
\end{array}
\right),\SSLL_{yy}=\left(
\begin{array}{ccccc}
 -\frac{2}{5} & 0 & 0 & 0 & 0 \\
 0 & -\frac{2}{5} & 0 & 0 & 0 \\
 0 & 0 & -\frac{2}{5} & 0 & 0 \\
 0 & 0 & 0 & \frac{8}{5} & 0 \\
 0 & 0 & 0 & 0 & -\frac{2}{5} \\
\end{array}
\right),\SSLL_{tt}=\left(
\begin{array}{ccccc}
 -\frac{2}{5} & 0 & 0 & 0 & 0 \\
 0 & -\frac{2}{5} & 0 & 0 & 0 \\
 0 & 0 & -\frac{2}{5} & 0 & 0 \\
 0 & 0 & 0 & -\frac{2}{5} & 0 \\
 0 & 0 & 0 & 0 & \frac{8}{5} \\
\end{array}
\right) \label{spkl5}
\end{align}
They also satisfy the condition $2\SSDD=-(\SSLL_{xx}+\SSLL_{yy}+\SSLL_{tt})$. The map to write the WL basis for $sl(5)$ theory in the terms of the PKLD matrices (\ref{pkl5}) and (\ref{spkl5}) is
\begin{footnotesize}
\besubeqs
\begin{align}
L_{-1}&\to   \text{SL}_{\text{--}}-\sqrt{\frac{3}{2}} L_{\text{xy}}- L_{\text{yt}}+\sqrt{\frac{3}{2}} P_x+\sqrt{\frac{3}{2}} \text{SL}_{\text{xy}}+ \text{SL}_{\text{yt}}+\sqrt{\frac{3}{2}} \text{SP}_x,\\
L_0&\to \frac{1}{2} \left(-D+\frac{7}{2} \text{SL}_{\text{tt}}+\frac{5}{2} \text{SL}_{\text{yy}}+\frac{3}{2} \text{SL}_{\text{xx}}\right),\\
L_1&\to - \text{SL}_{\text{++}}+\sqrt{\frac{3}{2}} K_x-\sqrt{\frac{3}{2}} L_{\text{xy}}- L_{\text{yt}}-\sqrt{\frac{3}{2}} \text{SK}_t-\sqrt{\frac{3}{2}} \text{SL}_{\text{xy}}- \text{SL}_{\text{yt}},
\end{align}
\esubeqs
\end{footnotesize}

\begin{footnotesize}
\besubeqs
\begin{align}
W^3{}_{-2}&\to 2 \left(\sqrt{6} K_x-\sqrt{6} L_{\text{xt}}+3 P_y+\sqrt{6} \text{SK}_x+\sqrt{6} \text{SL}_{\text{xt}}+3 \text{SP}_y\right),\\
W^3{}_{-1}&\to -3 \text{SL}_{\text{--}}-\sqrt{\frac{3}{2}} L_{\text{xy}}-3 L_{\text{yt}}-\sqrt{\frac{3}{2}} P_x+\sqrt{\frac{3}{2}} \text{SL}_{\text{xy}}+3 \text{SL}_{\text{yt}}-\sqrt{\frac{3}{2}} \text{SP}_x,\\
W^3{}_0&\to \frac{1}{2} \left(6 D+3 \text{SL}_{\text{tt}}-5 \text{SL}_{\text{xx}}-3 \text{SL}_{\text{yy}}\right),\\
W^3{}_1&\to  3\text{SL}_{\text{++}}-\sqrt{\frac{3}{2}} K_x-\sqrt{\frac{3}{2}} L_{\text{xy}}-3 L_{\text{yt}}+\sqrt{\frac{3}{2}} \text{SK}_x-\sqrt{\frac{3}{2}} \text{SL}_{\text{xy}}-3 \text{SL}_{\text{yt}},\\
W^3{}_2&\to 2 \left(-3 K_y+\sqrt{6} L_{\text{xt}}-\sqrt{6} P_x+3 \text{SK}_y+\sqrt{6} \left(\text{SL}_{\text{xt}}+\text{SP}_x\right)\right),\\
\end{align}
\esubeqs
\end{footnotesize}

\begin{footnotesize}
\besubeqs
\begin{align}
W^4{}_{-3}&\to 12 \left(K_y+P_t+\text{SK}_y+\text{SP}_t\right),\\
W^4{}_{-2}&\to -2 \sqrt{6}(K_x+L_{\text{xt}}+\text{SK}_x-\text{SL}_{\text{xt}}),\\
W^4{}_{-1}&\to \frac{4}{5} \left(3 \text{SL}_{\text{--}}+\sqrt{6} L_{\text{xy}}-3 L_{\text{yt}}-\sqrt{6} P_x-\sqrt{6} \text{SL}_{\text{xy}}+3 \text{SL}_{\text{yt}}-\sqrt{6} \text{SP}_x\right),\\
W^4{}_0&\to \frac{3}{5} \left(-6 D+ \text{SL}_{\text{tt}}-5 \text{SL}_{\text{yy}}-\text{SL}_{\text{xx}}\right),\\
W^4{}_1&\to \frac{4}{5} \left(-3 \text{SL}_{\text{++}}-\sqrt{6} K_x+\sqrt{6} L_{\text{xy}}-3 L_{\text{yt}}+\sqrt{6} \text{SK}_x+\sqrt{6} \text{SL}_{\text{xy}}-3 \text{SL}_{\text{yt}}\right),\\
W^4{}_2&\to 2 \sqrt{6}(L_{\text{xt}}+P_x+\text{SL}_{\text{xt}}-\text{SP}_x),\\
W^4{}_3&\to 12 \left(K_t+P_y-\text{SK}_t-\text{SP}_y\right),\\
\end{align}
\esubeqs
\end{footnotesize}

\begin{footnotesize}
\besubeqs
\begin{align}
W^5{}_{-4}&\to 24 \left(K_t+\text{SK}_t\right),\\
W^5{}_{-3}&\to 6 \left(-K_y+P_t-\text{SK}_y+\text{SP}_t\right),\\
W^5{}_{-2}&\to \frac{6}{7} \left(\sqrt{6} K_x-\sqrt{6} L_{\text{xt}}-4 P_y+\sqrt{6} \text{SK}_x+\sqrt{6} \text{SL}_{\text{xt}}-4 \text{SP}_y\right),\\
W^5{}_{-1}&\to \frac{6}{7} \left(- \text{SL}_{\text{--}}+\sqrt{6} L_{\text{xy}}-L_{\text{yt}}+\sqrt{6} P_x-\sqrt{6} \text{SL}_{\text{xy}}+\text{SL}_{\text{yt}}+\sqrt{6} \text{SP}_x\right),\\
W^5{}_0&\to \frac{6}{7} \left(2 D+ \text{SL}_{\text{tt}}+3 \text{SL}_{\text{xx}}- \text{SL}_{\text{yy}}\right),\\
W^5{}_1&\to \frac{6}{7} \left( \text{SL}_{\text{++}}+\sqrt{6} K_x+\sqrt{6} L_{\text{xy}}-L_{\text{yt}}-\sqrt{6} \text{SK}_x+\sqrt{6} \text{SL}_{\text{xy}}-\text{SL}_{\text{yt}}\right),\\
W^5{}_2&\to \frac{6}{7} \left(4 K_y+\sqrt{6} L_{\text{xt}}-\sqrt{6} P_x-4 \text{SK}_y+\sqrt{6} \left(\text{SL}_{\text{xt}}+\text{SP}_x\right)\right),\\
W^5{}_3&\to 6 \left(K_t-P_y-\text{SK}_t+\text{SP}_y\right),\\
W^5{}_4&\to 24 \left(\text{SP}_t-P_t\right).
\end{align}
\esubeqs
\end{footnotesize}

\subsection{Appendix E}
The group element that we can consider is 
 $ b=\rho \PPP_t-a_1 \PPP_y+\PPP_x+SL_{tt}$ and 2) $a_2 \PPP_t-a_1 \rho \PPP_y+\PPP_x+SL_{tt}$. 
Let us consider case 1) first. The obtained dreibein is 
\begin{align}
\left(
\begin{array}{ccc}
 e_{\rho}^{\tilde{x}} & e_{\rho}^{\tilde{y}} & e_{\rho}^{\tilde{t}} \\
 e_{\phi}^{\tilde{x}} & e_{\phi}^{\tilde{y}} & e_{\phi}^{\tilde{t}}\\
 e_{t}^{\tilde{x}}& e_t^{\tilde{y}} & e_t^{\tilde{t}} 
\end{array}
\right)
\end{align}
for $e_{\rho}^{\tilde{t}}=1$, $e_{\rho}^{\tilde{y}}=e_{\rho}^{\tilde{x}}=0$  and
\besubeqs
\begin{align}
  e_{i}^{\tilde{t}}&=\rho\left(-\frac{X^i}{2}+3 X^i_3+\frac{12 X^i_5}{7}-\frac{18 X^i_4}
  {5}\right) \\ 
  e_i^{\tilde{y}}&=\frac{1}{70} a_1 \left(35 X^i-210 X^i_3+252 X^i_4-120 X^i_5\right)\\
  e_i^{\tilde{x}}&=-\frac{\left(e^2-1\right) \left(35 \left(4 e^2-3\right) X^i-210 X^i_3+84 \left(1+2 e^2\right) X^i_4-120 X^i_5\right)}{140 e^2}.
  \end{align}\label{dreisl5}
  \esubeqs
  Here $i=\{\phi,t\}$, for $i=\phi$, $X^{\phi}=\mak, X^{\phi}_3=\mak_{0}^{(3)}, X^{\phi}_4=\mak_{0}^{(4)},X^{\phi}_5=\mak_{0}^{(5)}$ and for $i=t$, $X^t=\mu,X^t_3=\mu_3,X^t_4=\mu_4,X^t_5=\mu_5$. (Here we have also renamed the coordinates $x,y,t$ from the $SO(3,2)$ basis of generators from $x,y,t
\rightarrow\tilde{x},\tilde{y},\tilde{t}$). We set chemical potentials $\mu_3=\mu_4=0$ for simplicity and we obtain a metric with Ricci scalar $R=-\frac{2}{a_1^2}$. This is the only case in which we keep for now the dependency on $\phi$ of the fields and chemical potentials. Such metric can be easily transformed into the metric of the form 
\begin{align}
ds^2=dr^2+f_1(r)dt^2+f_2(r)d\phi dt+f_3(r)d\phi^2
\end{align}
for
\begin{footnotesize}
\besubeqs
\begin{align}
f_1(r)&=\frac{1}{4} e^{-2 (r+2)} \bigl(9 e^4 \left(\mak-\mak_{0}^{(3)}+\mak_{0}^{(4)}-\mak_{0}^{(5)}+\mu \right){}^2+4 e^{2 r+8} \left(6 (\mak+\mu )+\mak_{0}^{(4)}\right){}^2  \nonumber \\&-4 e^{2 r+6} \left(3 \left(7 (\mak+\mu )+\mak_{0}^{(5)}\right)+3 \mak_{0}^{(3)}+\mak_{0}^{(4)}\right) \left(6 (\mak+\mu )+\mak_{0}^{(4)}\right) \nonumber\\&+9 e^{4 r+4} \left(\mak-\mak_{0}^{(3)}+\mak_{0}^{(4)}-\mak_{0}^{(5)}+\mu \right){}^2+e^{2 r} \left(9 (\mak+\mu )+3 \mak_{0}^{(3)}-\mak_{0}^{(4)}+3 \mak_{0}^{(5)}\right){}^2\nonumber \\ 
 &-2 e^{2 r+2} \left(9 (\mak+\mu )+3 \mak_{0}^{(3)}-\mak_{0}^{(4)}+3 \mak_{0}^{(5)}\right) \left(3 \left(7 (\mak+\mu )+\mak_{0}^{(5)}\right)+3 \mak_{0}^{(3)}+\mak_{0}^{(4)}\right) \nonumber\\
 &+3 e^{2 r+4} \left(6 \mak_{0}^{(3)} \left(9 (\mak+\mu )-\mak_{0}^{(4)}+3 \mak_{0}^{(5)}\right)-6 \mak_{0}^{(4)} \left(\mak_{0}^{(5)}-5 (\mak+\mu )\right) \right. \nonumber\\ 
 & \left. +9 \left(25 (\mak+\mu )^2+6 \mak_{0}^{(5)} (\mak+\mu )+(\mak_{0}^{(5)}){}^2\right)+9 (\mak_{0}^{(3)}){}^2+5 (\mak_{0}^{(4)}){}^2\right)\bigr)\end{align} 
 \begin{align}
f_2(r)&=
\frac{1}{2} e^{-2(r+2)} \Big[
9 e^4 \left(-\mak+\mak_{0}^{(3)}-\mak_{0}^{(4)}+\mak_{0}^{(5)}\right)
\left(-\mak+\mak_{0}^{(3)}-\mak_{0}^{(4)}+\mak_{0}^{(5)}-\mu \right)
\nonumber \\ &
+4 e^{2r+8} \left(6 \mak+\mak_{0}^{(4)}\right)
\left(6 (\mak+\mu)+\mak_{0}^{(4)}\right)
\nonumber \\ &
+9 e^{4r+4} \left(-\mak+\mak_{0}^{(3)}-\mak_{0}^{(4)}+\mak_{0}^{(5)}\right)
\left(-\mak+\mak_{0}^{(3)}-\mak_{0}^{(4)}+\mak_{0}^{(5)}-\mu \right)
\nonumber \\ &
+e^{2r} \left(9 \mak+3 \mak_{0}^{(3)}-\mak_{0}^{(4)}+3 \mak_{0}^{(5)}\right)
\left(9 (\mak+\mu)+3 \mak_{0}^{(3)}-\mak_{0}^{(4)}+3 \mak_{0}^{(5)}\right)
\nonumber \\ &
+3 e^{2r+4} \Big(
3 \mak_{0}^{(3)} \left(18 \mak-2 \mak_{0}^{(4)}+6 \mak_{0}^{(5)}+9 \mu \right)
+\mak_{0}^{(4)} \left(15 (2 \mak+\mu)-6 \mak_{0}^{(5)}\right)
\nonumber \\ & \qquad
+9 \left(25 \mak (\mak+\mu)
+3 \mak_{0}^{(5)} (2 \mak+\mu)
+(\mak_{0}^{(5)})^2\right)
+9 (\mak_{0}^{(3)})^2
+5 (\mak_{0}^{(4)})^2
\Big)
\nonumber \\ &
-2 e^{2r+2} \Big(
9 \mak_{0}^{(3)} \left(5 (2 \mak+\mu)+2 \mak_{0}^{(5)}\right)
-6 \mak_{0}^{(4)} (2 \mak+\mu)
\nonumber \\ & \qquad
+9 \left(21 \mak (\mak+\mu)
+5 \mak_{0}^{(5)} (2 \mak+\mu)
+(\mak_{0}^{(5)})^2\right)
+9 (\mak_{0}^{(3)})^2
-(\mak_{0}^{(4)})^2
\Big)
\nonumber \\ &
-2 e^{2r+6} \Big(
9 \mu \left(28 \mak+2 \mak_{0}^{(3)}+3 \mak_{0}^{(4)}+2 \mak_{0}^{(5)}\right)
\nonumber \\ & \qquad
+2 \left(6 \mak+\mak_{0}^{(4)}\right)
\left(3 \left(7 \mak+\mak_{0}^{(5)}\right)
+3 \mak_{0}^{(3)}+\mak_{0}^{(4)}\right)
\Big)
\Big]
\end{align}
\esubeqs
\besubeqs
\begin{align}
f_3(r)&=\frac{1}{4} e^{-2 (r+2)} \bigl(9 e^4 \left(-\mak+\mak_{0}^{(3)}-\mak_{0}^{(4)}+\mak_{0}^{(5)}\right){}^2  \nonumber\\ &+3 e^{2 r+4} \left(225 \mak^2+6 \mak \left(9 \mak_{0}^{(3)}+5 \mak_{0}^{(4)}+9 \mak_{0}^{(5)}\right)+9 (\mak_{0}^{(3)}){}^2+5 (\mak_{0}^{(4)}){}^2+9 (\mak_{0}^{(5)}){}^2\right.\nonumber\\ & \left.-6 \mak_{0}^{(3)} \mak_{0}^{(4)}+18 \mak_{0}^{(3)} \mak_{0}^{(5)}
-6 \mak_{0}^{(4)} \mak_{0}^{(5)}\right)\nonumber\\ &+4 e^{2 r+8} \left(6 \mak+\mak_{0}^{(4)}\right){}^2 - 4 e^{2 r+6} \left(21 \mak+3 \mak_{0}^{(3)}+\mak_{0}^{(4)}+3 \mak_{0}^{(5)}\right) \left(6 \mak+\mak_{0}^{(4)}\right)\nonumber\\&+9 e^{4 r+4} \left(-\mak+\mak_{0}^{(3)}-\mak_{0}^{(4)}+\mak_{0}^{(5)}\right){}^2+e^{2 r} \left(9 \mak+3 \mak_{0}^{(3)}-\mak_{0}^{(4)}+3 \mak_{0}^{(5)}\right){}^2\nonumber\\&-2 e^{2 r+2} \left(9 \mak+3 \mak_{0}^{(3)}-\mak_{0}^{(4)}+3 \mak_{0}^{(5)}\right) \left(21 \mak+3 \mak_{0}^{(3)}+\mak_{0}^{(4)}+3 \mak_{0}^{(5)}\right)\bigr) 
\end{align}
\esubeqs
\end{footnotesize}
One can notice that due to the particular choice of the group element $b$, the fields are appearing in the metric as linear combinations. If there were dependency on  polynomials of $\rho$ in the group element, that would reflect in the metric with each field being multiplied by $\rho$ and there would be additional terms appearing due to projection procedure and transformation. Under our conditions, we see that requiring that the lienar combination of the fields takes certain value would impose constraint between the fields themselves. 
For the following step we set the fields and the chemical potentials to be constant. 
Transitioning to the coordinates $t\rightarrow x^++x^-$ and $\phi\rightarrow x^+-x^-$ leads to the metric which one can wirte as $ds^2=dr^2+g_1(r) dx^{+2}+g_2(r) dx^{+}dx^-+g_3(r)dx^{-2}$
\begin{footnotesize}
\besubeqs
\begin{align}
g_1(r)&=
\frac{1}{4} \bigr(16 e^4 \left(6 \mak+\mak_{0}^{(4)}+3 \mu \right){}^2-8 e^2 \left(42 \mak+6 \mak_{0}^{(3)}+2 \mak_{0}^{(4)}+6 \mak_{0}^{(5)}+21 \mu \right) \left(6 \mak+\mak_{0}^{(4)}+3 \mu \right)  \nonumber \\  &+\frac{\left(18 \mak+6 \mak_{0}^{(3)}-2 \mak_{0}^{(4)}+6 \mak_{0}^{(5)}+9 \mu \right){}^2}{e^4}+3 \left(225 (2 \mak+\mu )^2+60 \mak_{0}^{(4)} (2 \mak+\mu )\right. \nonumber \\   &\left.+108 \mak_{0}^{(5)} (2 \mak+\mu )+36 \mak_{0}^{(3)} \left(6 \mak+2 \mak_{0}^{(5)}+3 \mu \right)+36 (\mak_{0}^{(3)}){}^2+20 (\mak_{0}^{(4)}){}^2+36 (\mak_{0}^{(5)}){}^2-24 \mak_{0}^{(3)} \mak_{0}^{(4)}-24 \mak_{0}^{(4)} \mak_{0}^{(5)}\right)  \nonumber \\  &-\frac{2 \left(18 \mak+6 \mak_{0}^{(3)}-2 \mak_{0}^{(4)}+6 \mak_{0}^{(5)}+9 \mu \right) \left(42 \mak+6 \mak_{0}^{(3)}+2 \mak_{0}^{(4)}+6 \mak_{0}^{(5)}+21 \mu \right)}{e^2} \nonumber \\&+9 e^{-2 r} \left(2 \left(\mak+\mak_{0}^{(4)}-\mak_{0}^{(5)}\right)-2 \mak_{0}^{(3)}+\mu \right){}^2+9 e^{2 r} \left(2 \left(\mak+\mak_{0}^{(4)}-\mak_{0}^{(5)}\right)-2 \mak_{0}^{(3)}+\mu \right){}^2\bigl) \\
g_2(r)&=\frac{9 \mu ^2 \left(2 e^4 \cosh (2 r)+16 e^8-56 e^6+75 e^4-42 e^2+9\right)}{4 e^4}\\g_3(r)&=\frac{3}{2} \mu  \bigl(\frac{3 (18 \mak+6 \mak_{0}^{(3)}-2 \mak_{0}^{(4)}+6 \mak_{0}^{(5)}+9 \mu )}{e^4}-12 e^2 (2 \mak_{0}^{(3)}+3 \mak_{0}^{(4)}+2 (14 \mak+\mak_{0}^{(5)}+7 \mu )) \nonumber \\&-\frac{2 (63 (2 K+\mu )+30 \mak_{0}^{(3)}-4 \mak_{0}^{(4)}+30 \mak_{0}^{(5)})}{e^2} +e^{-2 r} (-6 \mak_{0}^{(3)}+6 (K+\mak_{0}^{(4)}-\mak_{0}^{(5)})+3 \mu )\nonumber  \\&+e^{2 r} (-6 \mak_{0}^{(3)}+6 (K+\mak_{0}^{(4)}-\mak_{0}^{(5)})+3 \mu )+54 \mak_{0}^{(3)}+16 e^4 (6 K+\mak_{0}^{(4)}+3 \mu )+30 \mak_{0}^{(4)}+9 (25 (2 K+\mu )+6 \mak_{0}^{(5)})\bigr)
\end{align}
\esubeqs
\end{footnotesize}
If we now choose that chemical potential takes the form $\mu=2(\mak_{0}^{(3)}-\mak_{0}^{(4)}+\mak_{0}^{(5)}-\mak)$ 
we obtain particulary simple form of the metric with $R=-2$. 
\begin{align}
ds^2&=dr^2+\frac{4}{e^4} \left(e^2-1\right)^4 \left(6 \mak_{0}^{(3)}-5 \mak_{0}^{(4)}+6 \mak_{0}^{(5)}\right){}^2 dx^{+2}\nonumber\\&+\frac{12}{e^4} \left(e^2-1\right)^3 \left(4 e^2-3\right) \left(6 \mak_{0}^{(3)}-5 \mak_{0}^{(4)}+6 \mak_{0}^{(5)}\right)\left(\mak-\mak_{0}^{(3)}+\mak_{0}^{(4)}-\mak_{0}^{(5)}\right) dx^+ dx^-\nonumber\\&+\frac{9}{e^4}  \left(\mak-\mak_{0}^{(3)}+\mak_{0}^{(4)}-\mak_{0}^{(5)}\right){}^2 \left(2 e^4 cosh (2 r)+16 e^8-56 e^6+75 e^4-42 e^2+9\right)dx^{-2}
\end{align}
The metric for the group element 2) leads to dreibein which of similar form as the dreibein for the group element 1). The difference is that this time $e_i^{\tilde{x}}$ contains $\rho$ while $e_i^{\tilde{y}}$ does not. When we write a metric $ds^2=\frac{1}{f_{rr}}dr^2+f_{tt}dt^2+f_{t\phi}dtd\phi+r^2d\phi^2$ for this dreibein we obtain 
\begin{footnotesize}
    \besubeqs
    \begin{align}
    f_{rr}&=\Big[\left(\left(e^2-1\right)^2 \left(12 e^2 \mak-9 \mak-3 \mak_{0}^{(3)}+2 e^2 \mak_{0}^{(4)}+\mak_{0}^{(4)}-3 \mak_{0}^{(5)}\right){}^2-e^4 r^2\right) \left(\mak_{0}^{(3)} \left(18 \left(4 e^4 a_2^2+\left(e^2-1\right)^2\right) \mak_{0}^{(5)}\right.\right.\nonumber\\&\left.\left.-e_2 \mak_{0}^{(4)}-6 e_3 \mak\right)-6 \mak_{0}^{(5)} \left(\left(12 e^4 a_2^2+2 e^6-3 e^4+1\right) \mak_{0}^{(4)}+\text{e3} \mak\right)+\mak_{0}^{(4)} \left(\left(36 e^4 a_2^2+\left(1+e^2-2 e^4\right)^2\right) \mak_{0}^{(4)}+6 e_4 \mak\right)\nonumber \right.\\& \left.+9 \left(4 e^4 a_2^2+\left(e^2-1\right)^2\right) (\mak_{0}^{(3)}){}^2+9 \left(4 e^4 a_2^2+\left(e^2-1\right)^2\right) (\mak_{0}^{(5)}){}^2+e_1 \mak^2-e^4 r^2\right)\Big]\frac{1}{e^8 a_2^2 r^2}\\
     f_{tt}&=\frac{1}{36} \left(\frac{\mu ^2 r^2}{\left(\mak-\mak_{0}^{(3)}+\mak_{0}^{(4)}-\mak_{0}^{(5)}\right){}^2}+\frac{2 \mu  r^2}{\mak-\mak_{0}^{(3)}+\mak_{0}^{(4)}-\mak_{0}^{(5)}}+r^2\right)\nonumber \\&-\frac{\left(e^2-1\right)^3 \left(6 \mak_{0}^{(3)}-5 \mak_{0}^{(4)}+6 \mak_{0}^{(5)}\right)}{9 e^4 \left(\mak-\mak_{0}^{(3)}+\mak_{0}^{(4)}-\mak_{0}^{(5)}\right){}^2} \mu  \left(\mak_{0}^{(3)} \left(-3 \left(2 e^2-1\right) (2 \mak+\mu )-2 \left(2+e^2\right) \mak_{0}^{(4)}+6 \mak_{0}^{(5)}\right)   \right. \nonumber  \\&\left.+3 \mak_{0}^{(5)} \left(\mak_{0}^{(5)}-\left(2 e^2-1\right) (2 \mak+\mu )\right)+\mak_{0}^{(4)} \left(\left(7 e^2-4\right) (2 \mak+\mu )-2 \left(2+e^2\right) \mak_{0}^{(5)}\right)+3 \left(4 e^2-3\right) \mak (\mak+\mu ) \right. \nonumber \\ & \left. +3 (\mak_{0}^{(3)}){}^2+\left(1+2 e^2\right) (\mak_{0}^{(4)}){}^2\right)\\
f_{t\phi}&=\bigl[-2 \left(6 \mak_{0}^{(3)}-5 \mak_{0}^{(4)}+6 \mak_{0}^{(5)}\right) \mu  \left(9 \mak+3 \mak_{0}^{(3)}-\mak_{0}^{(4)}+3 \mak_{0}^{(5)}\right) 
 \nonumber \\ & +2 e^6 \left(6 \mak_{0}^{(3)}-5 \mak_{0}^{(4)}+6 \mak_{0}^{(5)}\right) \mu  \left(45 \mak+3 \mak_{0}^{(3)}+5 \mak_{0}^{(4)}+3 \mak_{0}^{(5)}\right)\nonumber\\&+2 e^2 \left(6 \mak_{0}^{(3)}-5 \mak_{0}^{(4)}+6 \mak_{0}^{(5)}\right) \mu  \left(39 \mak+9 \mak_{0}^{(3)}-\mak_{0}^{(4)}+9 \mak_{0}^{(5)}\right)+4 e^8 \left(-6 \mak_{0}^{(3)}+5 \mak_{0}^{(4)}-6 \mak_{0}^{(5)}\right) \mu  \left(6 \mak+\mak_{0}^{(4)}\right)\nonumber\\&+e^4 \left(\mu  \left(r^2-6 \left(6 \mak_{0}^{(3)}-5 \mak_{0}^{(4)}+6 \mak_{0}^{(5)}\right) \left(3 \left(7 \mak+\mak_{0}^{(5)}\right)+3 \mak_{0}^{(3)}+\mak_{0}^{(4)}\right)\right) \right.\nonumber\\&\left.+r^2 \left(\mak-\mak_{0}^{(3)}+\mak_{0}^{(4)}-\mak_{0}^{(5)}\right)\right)\bigr]\frac{1}{3 e^4 \left(\mak-\mak_{0}^{(3)}+\mak_{0}^{(4)}-\mak_{0}^{(5)}\right)}
\end{align}
\esubeqs
\end{footnotesize}
Here the coefficients are $e_1=9 \left(e^4 \left(4 a_2^2+73\right)+16 e^8-56 e^6-42 e^2+9\right)$, \\ $e_2=6 \left(3 e^4 \left(4 a_2^2-1\right)+2 e^6+1\right)$, $e_3=3 \left(e^4 \left(4 a_2^2-11\right)+4 e^6+10 e^2-3\right)$, $e_4=3 e^4 \left(4 a_2^2+3\right)+8 e^8-18 e^6+4 e^2-3$.
When we turn off the fields $\mak_{0}^{(4)}=\mak_{0}^{(5)}=\mak=0$ the metric reduces to 
\begin{footnotesize}
    \besubeqs
\begin{align}
ds^2&=\frac{16 e^8 a_2^2 r^2}{\left(\left(e^2-1\right)^2 (\mak_{0}^{(3)}){}^2-4 e^4 r^2\right) \left(\left(e^4 \left(4 a_2^2+1\right)-2 e^2+1\right) (\mak_{0}^{(3)}){}^2-4 e^4 r^2\right)} dr^2\\&+ \left(r^2 \left(2-\frac{2 \mu}{\mak_{0}^{(3)}}\right)-\frac{2 \left(e^2-1\right)^3 \mak_{0}^{(3)} x}{e^4}\right)dtd\phi\\&+\left(\frac{r^2 (\mak_{0}^{(3)}-\mu)^2}{(\mak_{0}^{(3)}){}^2}-\frac{2 \left(e^2-1\right)^3 \mu \left(\mak_{0}^{(3)}-2 e^2 \mu+\mu\right)}{e^4}\right)dt^2+r^2d\phi^2
\end{align}
\esubeqs
\end{footnotesize}

Which can after coordinate transformations, rescaling and absorbing the parameters be brought to a form 
\begin{align}
ds^2=\frac{dr^2}{(\mak_{0}^{(3)}){}^2+r^2}+2 \mak_{0}^{(3)} dt d\phi +r^2 d\phi ^2-dt^2. \label{sl5lob}
\end{align}


From (\ref{sl5lob}) we recognize the Lobachevsky form. 

The embedding of PKLD basis in WL basis is not the same as in the gauge theories based on the $sl(4)$ and the $so(3,2)$ algebra. That results with the metric that can be reduced to Lobachevsky form for the group element $b=a_2 \PPP_t-a_1\rho \PPP_y+\PPP_x+SL_{tt}$. For the  $SO(3,2)$ group, the element
that led to Lobachevsky metric was $b=\PPP_y+\DDD+c(\rho)\PPP_x$. 
Another difference due to embedding is that in $sl(5)$ one needs group element written in terms of all generators of translations for dreibein not to vanish.

\footnotesize
\bibliographystyle{unsrt}
\bibliography{megabib.bib}

\begin{thebibliography}{10}

\bibitem{Fradkin:1985am}
E.~S. Fradkin and Arkady~A. Tseytlin.
\newblock {Conformal Supergravity}.
\newblock {\em Phys. Rept.}, 119:233--362, 1985.

\bibitem{Pope:1989vj}
C.~N. Pope and P.~K. Townsend.
\newblock {Conformal Higher Spin in (2+1)-dimensions}.
\newblock {\em Phys. Lett.}, B225:245--250, 1989.

\bibitem{Fradkin:1989xt}
E.~S. Fradkin and V.~{\relax Ya}. Linetsky.
\newblock {A Superconformal Theory of Massless Higher Spin Fields in $D$ =
  (2+1)}.
\newblock {\em Mod. Phys. Lett.}, A4:731, 1989.
\newblock [Annals Phys.198,293(1990)].

\bibitem{Campoleoni:2010zq}
Andrea Campoleoni, Stefan Fredenhagen, Stefan Pfenninger, and Stefan Theisen.
\newblock {Asymptotic symmetries of three-dimensional gravity coupled to
  higher-spin fields}.
\newblock {\em JHEP}, 1011:007, 2010.

\bibitem{Campoleoni:2008jq}
Andrea Campoleoni, Dario Francia, Jihad Mourad, and Augusto Sagnotti.
\newblock {Unconstrained Higher Spins of Mixed Symmetry. I. Bose Fields}.
\newblock {\em Nucl. Phys.}, B815:289--367, 2009.

\bibitem{Grigoriev:2019xmp}
Maxim Grigoriev, Iva Lovrekovic, and Evgeny Skvortsov.
\newblock {New Conformal Higher Spin Gravities in $3d$}.
\newblock {\em JHEP}, 01:059, 2020.

\bibitem{Henneaux:2010xg}
Marc Henneaux and Soo-Jong Rey.
\newblock {Nonlinear $W_{infinity}$ as Asymptotic Symmetry of Three-Dimensional
  Higher Spin Anti-de Sitter Gravity}.
\newblock {\em JHEP}, 1012:007, 2010.

\bibitem{Brown:1986nw}
J.~David Brown and M.~Henneaux.
\newblock {Central Charges in the Canonical Realization of Asymptotic
  Symmetries: An Example from Three-Dimensional Gravity}.
\newblock {\em Commun. Math. Phys.}, 104:207--226, 1986.

\bibitem{Bergshoeff:1989ns}
E.~Bergshoeff, M.~P. Blencowe, and K.~S. Stelle.
\newblock {Area Preserving Diffeomorphisms and Higher Spin Algebra}.
\newblock {\em Commun. Math. Phys.}, 128:213, 1990.

\bibitem{Blencowe:1988gj}
M.P. Blencowe.
\newblock {A Consistent Interacting Massless Higher Spin Field Theory in $D$ =
  (2+1)}.
\newblock {\em Class.Quant.Grav.}, 6:443, 1989.

\bibitem{Campoleoni:2011hg}
Andrea Campoleoni, Stefan Fredenhagen, and Stefan Pfenninger.
\newblock {Asymptotic W-symmetries in three-dimensional higher-spin gauge
  theories}.
\newblock {\em JHEP}, 09:113, 2011.

\bibitem{Grumiller:2016pqb}
Daniel Grumiller and Max Riegler.
\newblock {Most general AdS$_{3}$ boundary conditions}.
\newblock {\em JHEP}, 10:023, 2016.

\bibitem{Afshar:2016wfy}
Hamid Afshar, Stephane Detournay, Daniel Grumiller, Wout Merbis, Alfredo Perez,
  David Tempo, and Ricardo Troncoso.
\newblock {Soft Heisenberg hair on black holes in three dimensions}.
\newblock {\em Phys. Rev. D}, 93(10):101503, 2016.

\bibitem{Hawking:2016msc}
Stephen~W. Hawking, Malcolm~J. Perry, and Andrew Strominger.
\newblock {Soft Hair on Black Holes}.
\newblock {\em Phys. Rev. Lett.}, 116(23):231301, 2016.

\bibitem{Lovrekovic:2021dvi}
Iva Lovrekovic.
\newblock {Conformal Carrollian spin-3 gravity in 3D}.
\newblock {\em Phys. Rev. D}, 105(12):124065, 2022.

\bibitem{Lovrekovic:2022lwv}
I.~Lovrekovic and K.~Schaefer.
\newblock {Conformal Galilean Spin-3 Gravity in 3d}.
\newblock 12 2022.

\bibitem{Horne:1988jf}
James~H. Horne and Edward Witten.
\newblock {Conformal Gravity in Three-dimensions as a Gauge Theory}.
\newblock {\em Phys. Rev. Lett.}, 62:501--504, 1989.

\bibitem{Bekaert:2013zya}
Xavier Bekaert and Maxim Grigoriev.
\newblock {Higher order singletons, partially massless fields and their
  boundary values in the ambient approach}.
\newblock {\em Nucl. Phys.}, B876:667--714, 2013.

\bibitem{Brust:2016gjy}
Christopher Brust and Kurt Hinterbichler.
\newblock {Free $\square^{k}$ scalar conformal field theory}.
\newblock {\em JHEP}, 02:066, 2017.

\bibitem{Banados:1994tn}
Maximo Banados.
\newblock {Global charges in Chern-Simons field theory and the (2+1) black
  hole}.
\newblock {\em Phys. Rev. D}, 52:5816--5825, 1996.

\bibitem{Banados:1998ta}
Maximo Banados, Thorsten Brotz, and Miguel~E. Ortiz.
\newblock {Boundary dynamics and the statistical mechanics of the
  (2+1)-dimensional black hole}.
\newblock {\em Nucl. Phys. B}, 545:340--370, 1999.

\bibitem{Banados:1998gg}
Maximo Banados.
\newblock {Three-dimensional quantum geometry and black holes}.
\newblock {\em AIP Conf. Proc.}, 484(1):147--169, 1999.

\bibitem{Carlip:2005zn}
Steven Carlip.
\newblock {Conformal field theory, (2+1)-dimensional gravity, and the BTZ black
  hole}.
\newblock {\em Class. Quant. Grav.}, 22:R85--R124, 2005.

\bibitem{Grumiller:2016kcp}
Daniel Grumiller, Alfredo Perez, Stefan Prohazka, David Tempo, and Ricardo
  Troncoso.
\newblock {Higher Spin Black Holes with Soft Hair}.
\newblock {\em JHEP}, 10:119, 2016.

\bibitem{Campoleoni:2024ced}
Andrea Campoleoni and Stefan Fredenhagen.
\newblock {Higher-Spin Gauge Theories in Three Spacetime Dimensions}.
\newblock {\em Lect. Notes Phys.}, 1028:121--267, 2024.

\bibitem{Eastwood2008}
Michael Eastwood and Thomas Leistner.
\newblock {\em Higher Symmetries of the Square of the Laplacian}, pages
  319--338.
\newblock Springer New York, New York, NY, 2008.

\bibitem{Joung:2014qya}
Euihun Joung and Karapet Mkrtchyan.
\newblock {Notes on higher-spin algebras: minimal representations and structure
  constants}.
\newblock {\em JHEP}, 05:103, 2014.

\bibitem{Perez:2012cf}
Alfredo Perez, David Tempo, and Ricardo Troncoso.
\newblock {Higher spin gravity in 3D: Black holes, global charges and
  thermodynamics}.
\newblock {\em Phys. Lett. B}, 726:444--449, 2013.

\bibitem{Perez:2013xi}
Alfredo Perez, David Tempo, and Ricardo Troncoso.
\newblock {Higher spin black hole entropy in three dimensions}.
\newblock {\em JHEP}, 04:143, 2013.

\bibitem{deBoer:2013gz}
Jan de~Boer and Juan~I. Jottar.
\newblock {Thermodynamics of higher spin black holes in $AdS_3$}.
\newblock {\em JHEP}, 01:023, 2014.

\bibitem{Bunster:2014mua}
Claudio Bunster, Marc Henneaux, Alfredo Perez, David Tempo, and Ricardo
  Troncoso.
\newblock {Generalized Black Holes in Three-dimensional Spacetime}.
\newblock {\em JHEP}, 05:031, 2014.

\bibitem{Lovrekovic:2025dwd}
I.~Lovrekovic.
\newblock {A note on one-parameter subgroups of SO(3,2)}.
\newblock 12 2025.

\bibitem{Banados:1992gq}
Maximo Banados, Marc Henneaux, Claudio Teitelboim, and Jorge Zanelli.
\newblock {Geometry of the (2+1) black hole}.
\newblock {\em Phys. Rev. D}, 48:1506--1525, 1993.
\newblock [Erratum: Phys.Rev.D 88, 069902 (2013)].

\bibitem{Oliva:2009hz}
Julio Oliva, David Tempo, and Ricardo Troncoso.
\newblock {Static spherically symmetric solutions for conformal gravity in
  three dimensions}.
\newblock {\em Int. J. Mod. Phys. A}, 24:1588--1592, 2009.

\bibitem{Mannheim:1988dj}
Philip~D. Mannheim and Demosthenes Kazanas.
\newblock {Exact Vacuum Solution to Conformal Weyl Gravity and Galactic
  Rotation Curves}.
\newblock {\em Astrophys. J.}, 342:635--638, 1989.

\bibitem{Riegert:1984zz}
Ronald~J. Riegert.
\newblock {Birkhoff's Theorem in Conformal Gravity}.
\newblock {\em Phys. Rev. Lett.}, 53:315--318, 1984.

\bibitem{Emparan:2022ijy}
Roberto Emparan, Juan~F. Pedraza, Andrew Svesko, Marija Toma\v{s}evi\'c, and
  Manus~R. Visser.
\newblock {Black holes in dS$_{3}$}.
\newblock {\em JHEP}, 11:073, 2022.

\bibitem{Bertin:2012qw}
Mario Bertin, Sabine Ertl, Hossein Ghorbani, Daniel Grumiller, Niklas
  Johansson, and Dmitri Vassilevich.
\newblock {Lobachevsky holography in conformal Chern-Simons gravity}.
\newblock {\em JHEP}, 06:015, 2013.

\end{thebibliography}

\end{document}